\documentclass[prc,aps,twocolumn,showpacs,amssymb,superscriptaddress,fleqn]{revtex4}

\usepackage{amsmath}
\usepackage{color}
\usepackage{graphicx}
\usepackage{dcolumn}
\usepackage{mathtools}
\usepackage{relsize}
\usepackage{float}
\usepackage{braket}
\newcommand{\vect}{\boldsymbol}
\DeclareSymbolFont{largesymbol}{OMX}{yhex}{m}{n}
\DeclareMathAccent{\Widehat}{\mathord}{largesymbol}{"62}

\begin{document}

\title{Realistic shell-model calculations for $p$-shell nuclei \\
  including contributions of a chiral three-body force}

\author{T. Fukui}
\affiliation{Istituto Nazionale di Fisica Nucleare, \\
Complesso Universitario di Monte  S. Angelo, Via Cintia, I-80126 Napoli, Italy}
\author{L. De Angelis}
\affiliation{Istituto Nazionale di Fisica Nucleare, \\
Complesso Universitario di Monte  S. Angelo, Via Cintia, I-80126 Napoli, Italy}
\author{Y. Z. Ma}
\affiliation{School of Physics, and State Key Laboratory of Nuclear
  Physics and Technology, \\
Peking University, Beijing I-100871, China}
\author{L. Coraggio}
\affiliation{Istituto Nazionale di Fisica Nucleare, \\
Complesso Universitario di Monte  S. Angelo, Via Cintia, I-80126 Napoli, Italy}
\author{A. Gargano}
\affiliation{Istituto Nazionale di Fisica Nucleare, \\
Complesso Universitario di Monte  S. Angelo, Via Cintia, I-80126 Napoli, Italy}
\author{N. Itaco}
\affiliation{Istituto Nazionale di Fisica Nucleare, \\ 
Complesso Universitario di Monte  S. Angelo, Via Cintia, I-80126 Napoli, Italy}
\affiliation{Dipartimento di Matematica e Fisica, Universit\`a degli
  Studi della Campania ``Luigi Vanvitelli'', viale Abramo Lincoln 5 -
  I-81100 Caserta, Italy}
\author{F. R. Xu}
\affiliation{School of Physics, and State Key Laboratory of Nuclear
  Physics and Technology, \\
Peking University, Beijing I-100871, China}
 
\begin{abstract}
In this paper we present an evolution of our derivation of the
shell-model effective Hamiltonian, namely introducing effects of
three-body contributions.
More precisely, we consider a three-body potential at
next-to-next-to-leading order in chiral perturbation theory, and the
induced three-body forces that arise from many-body correlations among
valence nucleons.
The first one is included, in the derivation of the effective
Hamiltonian for one- and two-valence nucleon-systems, at first order
in the many-body perturbation theory.
Namely, we include only the three-body interaction between one
or two valence nucleons and those belonging to the core.
For nuclei with more than two valence particles, both induced - turned
on by the two-body potential - and genuine three-body forces come
into play. 
Since it is difficult to perform shell-model calculations with
three-body forces, these contributions are estimated for the
ground-state energy only.
In order to establish the reliability of our approximations, we focus
attention on nuclei belonging to the $p$ shell, aiming to benchmark
our calculations against those performed with the {\it ab initio}
no-core shell-model.
The obtained results are satisfactory, and pave the way to the
application of our approach to nuclear systems with heavier masses.
\end{abstract}


\maketitle

\section{Introduction}
\label{intro}
The shell model (SM) is a fundamental tool for the microscopic
description of nuclear structure, and its most appealing feature is to
reduce the complexity of a many-body problem, where the degrees of
freedom of all the individual nucleons are explicitly taken into account,
to the one where only the valence nucleons interact in a limited model
space.

Within this framework, it is highly desirable to derive the SM
parameters, namely the single-particle (SP) energies and the two-body
matrix elements (TBME) of the residual interaction from realistic
nuclear forces.
This approach is the so-called realistic shell model (RSM), and its
roots trace back to the seminal paper by Kuo and Brown \cite{Kuo66}
more than fifty years ago, where a SM effective Hamiltonian
$H_{\rm eff}$ for $sd$-shell nuclei was derived starting from the
hard-core Hamada-Johnston potential \cite{Hamada62}.
Some historical developments of RSM may be found in review papers
\cite{Hjorth95,Coraggio09a}, and a certain number of fundamental
papers on this topic are collected in Ref. \cite{Brown10b}.

Our approach to derive $H_{\rm eff}$ is based on the
energy-independent linked-diagram perturbation theory \cite{Kuo90},
where the pivotal role is played by the perturbative expansion of the
$\hat{Q}$-box vertex function, that is a collection of irreducible
valence-linked Goldstone diagrams.
The effective Hamiltonian is obtained solving iteratively non-linear matrix
equations, that are expressed in terms of the $\hat{Q}$-box \cite{Suzuki80}.

Recently, an alternative way to derive $H_{\rm eff}$, framed within a
non-perturbative scheme, has been proposed \cite{Bogner14}.
This approach is an application of the in-medium similarity
renormalization group \cite{Hergert16}, and may provide a new
and valuable tool for the development of the RSM.

In a previous paper \cite{Coraggio12a}, we have described in detail the
process to derive $H_{\rm eff}$, and the procedures we apply to check
both the convergence properties of the perturbative expansion and the
weak dependence of the shell-model results upon the harmonic
oscillator (HO) parameter $\hbar \omega$.
The latter dependence is introduced, as in all many-body techniques
employing the HO auxiliary potential, by the
truncation of the number of intermediate states in the sum of the
perturbative expansion.

Moreover, to check the validity of our approach, we have performed
benchmark calculations comparing the outcome of the diagonalization of
RSM Hamiltonians with that of an {\it ab initio} method, such as the
no-core shell model (NCSM) \cite{Navratil04,Navratil07a}.
To this end, we derived $p$-shell effective Hamiltonians starting from a
realistic nuclear potential based on the chiral perturbation theory (ChPT)
at next-to-next-to-next-to-leading order (N$^3$LO) \cite{Entem02}, but
taking into account only the two-body ($2N$) component of this potential.

The comparison between the results obtained is very satisfactory,
especially considering that in NCSM the degrees of freedom of
all constituent nucleons are taken into account, while in RSM the
eigenfunctions contain explicitly configurations of the valence
nucleons only, that are constrained to a model space limited to the
$0p_{3/2}$ and $0p_{1/2}$ orbitals.

As a matter of fact, the low-lying energy spectra of some $p$-shell
nuclei calculated with RSM nicely agree with those by NCSM, while the
discrepancy of the calculated ground-state energies, with respect to
the $^4$He core, grows with the mass number $A$.
This can be explained by bearing in mind that our $H_{\rm
  eff}$ is derived for one- and two-valence nucleon-systems, while it
neglects the many-body ($>2$) components of $H_{\rm eff}$, that arise
from the interaction via the two-body force of the many-valence
nucleons with core excitations as well as with virtual intermediate
nucleons scattered above the model space.

In the present work, we address this issue, by calculating
the effect on the ground-state (g.s.) energies of three-body
correlation diagrams \cite{Ellis77,Polls83}, and also including in our
$H_{\rm eff}$, aside the chiral N$^3$LO two-body potential
\cite{Entem02,Machleidt11}, a chiral N$^2$LO three-body potential
\cite{Navratil07a} whose effects are considered at first-order in
perturbation theory.

So far, modern nuclear structure calculations have evidenced the role
played by three-nucleon ($3N$) forces, in particular for light nuclei
with $A \leq 12$ (see, for example, Refs. \cite{Pieper01,Pieper05}).
Our goal is to obtain an improvement of the reproduction of the
spectroscopic properties of $p$-shell nuclei, and benchmark our
results against those in Refs. \cite{Navratil07a,Maris13}, by including
the same chiral three-body potential.

It is worth mentioning that our approach to treat the microscopic $3N$
potential is similar to that in
Refs. \cite{Otsuka10,Holt13a,Holt14,Simonis16} where, aside a
realistic two-body low-momentum potential, only first-order
contributions of the normal-ordered two-body parts of $3N$ forces have
been taken explicitly into account.

The paper is organized as follows. 
In Section \ref{outline} we give an outline of the derivation of our
shell-model effective Hamiltonian within a perturbative approach, and of
our procedure to include three-body effects.
Section \ref{results} is devoted to compare our RSM results with those
provided by the {\em ab initio} NCSM \cite{Navratil07a,Maris13}. 
Concluding remarks and outlook of our future commitments are given in
Section \ref{conclusions}.
In Appendix, details of the calculations of the matrix elements of
the N$^2$LO three-body potential are reported.

\section{Theoretical framework}
\label{outline}
As mentioned in the Introduction, a detailed description of the
procedure we apply to derive $H_{\rm eff}$ within the many-body
perturbation theory has been reported in Ref. \cite{Coraggio12a}. 

We start our calculations by considering a high-precision
nucleon-nucleon ($NN$) potential derived within the ChPT at
next-to-next-to-next-to-leading order \cite{Entem02,Machleidt11}.
In the chiral perturbative expansion the $3N$ potentials appear from
N$^2$LO on, and we consider also its contributions in the derivation of
the $H_{\rm eff}$.

\begin{figure}[h]
\begin{center}
\includegraphics[scale=0.60,angle=0]{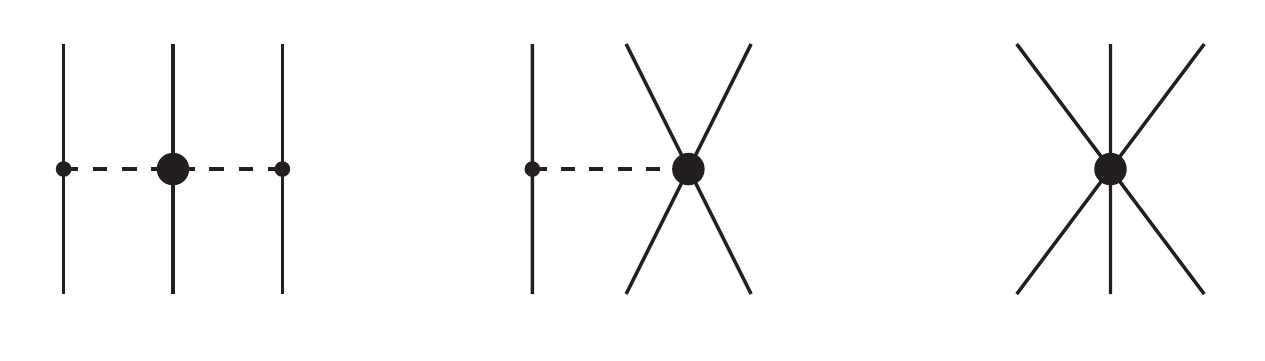}
\caption{The three-nucleon potential at N$^2$LO.
From left to right: 2$\pi$-exchange, 1$\pi$-exchange, and contact
diagrams.}
\label{n2lo3f}
\end{center}
\end{figure}

This $3N$ potential consists of three components (see
Fig. \ref{n2lo3f}), namely the two-pion ($2\pi$) exchange term
$V_{3N}^{(2\pi)}$, the one-pion ($1\pi$) exchange plus contact term
$V_{3N}^{(1\pi)}$, and the contact term $V_{3N}^{\textrm{(ct)}}$.

It should be pointed out that the low-energy constants (LECs)
$c_1$, $c_3$, and $c_4$, appearing in $V_{3N}^{(2\pi)}$, are
the same as those in the $NN$ potential, so their values are fixed by
the renormalization procedure that is performed for the two-body
N$^3$LO potential \cite{Machleidt11}.
However, the $3N$ 1$\pi$-exchange term and the contact interaction are
characterized by two extra LECs (known as $c_D$ and $c_E$,
respectively), which cannot be constrained by two-body observables,
and need to be fitted in order to reproduce observables in systems
with mass $A>2$.

Since we intend to benchmark our SM calculations against those in Refs.
\cite{Navratil07a,Maris13}, in this work we adopt $c_D=-1$, as
reported in Ref. \cite{Navratil07a}, and $c_E=-0.34$, as may be
inferred from Fig. 1 in the same reference.

The N$^2$LO $3N$ potential is defined in momentum space and, in order
to employ it to derive a shell-model effective interaction, we have
calculated its matrix elements in the HO basis following a procedure
similar to that indicated in Ref. \cite{Navratil07b}.
Actually, there is a difference about the calculation of the two-pion
exchange term with our formalism and the one reported in
Ref. \cite{Navratil07b}, and the details of our calculations are
reported in Appendix.

Note that the Coulomb potential is explicitly taken into account in
our calculations.

After choosing the $NN$ and $3N$ potentials, our following step is to
derive a SM effective Hamiltonian for one- and two-valence nucleon
systems within a model space spanned by the two proton and neutron
orbitals $0p_{3/2}$ and $0p_{1/2}$, outside the doubly-closed $^{4}$He core.

To this end, an auxiliary one-body potential $U$ is introduced in
order to break up the intrinsic Hamiltonian for a system of $A$
nucleons as the sum of a one-body term $H_0$, which describes the
independent motion of the nucleons, and a residual interaction $H_1$:

\begin{align}
\label{smham_int}
&H_{int} =  ~ \nonumber\\ 
&\left ( 1 - \frac{1}{A} \right ) \sum_{i} \frac{p_i^2}{2m} + \sum_{i
  < j} \left ( V^{NN}_{ij}- \frac{{\bf p_i} \cdot {\bf p_j}}{mA}
\right ) +  \nonumber\\ 
& \sum_{i < j < k} V^{3N}_{ijk} = \left[ \sum_{i} (\frac{p_i^2}{2m} + U_i) \right] +
\left[ \sum_{i < j} (V^{NN}_{ij} - \right.     ~ \nonumber\\
&\left. U_i - \frac{p^2_i}{2mA} -  \frac{\bf p_i \cdot p_j}{mA}) +
  \sum_{i < j < k} V^{3N}_{ijk} \right] = H_0+H_1 ~~, 
\end{align}

\noindent
where $i,j,k$ indices run from 1 to the mass number $A$, and ${\bf p}$
is the momentum of the nucleon.
Note that, in order to compare RSM with NCSM results, we
have to employ a purely intrinsic Hamiltonian by removing the
center-of-mass (CM) kinetic energy.
This introduces a dependence on the mass number $A$ that is relevant
for light systems, such as those belonging to the $p$ shell, but that
is strongly suppressed for heavier nuclei.

The diagonalization of the many-body Hamiltonian $H_{int}$ in an
infinite Hilbert space is unfeasible, and our eigenvalue problem is
then reduced to the one for an effective Hamiltonian $H_{\rm eff}$ in
a truncated model space.
Since $H_{int}$ has been broken up into two terms, we  define the
reduced model space in terms of a finite subset of $H_0$ eigenvectors.
In our calculation we choose as auxiliary potential $U$ the HO potential.

In this paper, we resort to the Kuo-Lee-Ratcliff (KLR) folded-diagram
expansion \cite{Kuo71,Kuo90} to calculate $H_{\rm eff}$, and this can
be done by way of a perturbative expansion of the vertex function
$\hat{Q}$-box

\begin{equation}
\hat{Q} (\epsilon) = P H_1 P + P H_1 Q \frac{1}{\epsilon - Q H Q} Q
H_1 P ~~,\label{qbox}
\end{equation}

\noindent
as defined in Ref. \cite{Suzuki80}.

In our calculations we expand the $\hat{Q}$-box in terms of
irreducible valence-linked one- and two-body Goldstone diagrams
through third order in $H_1$, for contributions with $2N$ vertices
\cite{Coraggio12a}, and up to first order for those with a
$3N$ vertex.
Then, to have a better estimate of the value to which the perturbation
series should converge, we resort to the Pad\'e approximant theory
\cite{Baker70,Ayoub79}, and calculate the Pad\'e approximant $[2|1]$
of the $\hat{Q}$-box, as suggested in \cite{Hoffmann76}:

\begin{equation}
[2|1] = V_{Qbox}^0 + V_{Qbox}^1 +V_{Qbox}^2
(1-(V_{Qbox}^2)^{-1}V_{Qbox}^{3})^{-1}~~.
\label{padeq}
\end{equation}

The $V_{Qbox}^n$ is the square non-singular matrix representing the
$n$th-order contribution to the $\hat{Q}$-box in the perturbative
expansion.

We have reviewed the calculation of our SM effective Hamiltonian
$H_{\rm eff}$ in Ref. \cite{Coraggio12a}, where details of the
diagrammatic expansion of the $\hat{Q}$-box and its perturbative
properties are also reported.

In terms of the $\hat{Q}$-box, the SM effective Hamiltonian can be
written in an operator form as

\begin{equation}
H_{\rm eff} = \hat{Q} - \hat{Q'} \int \hat{Q} + \hat{Q'} \int \hat{Q} \int
\hat{Q} - \hat{Q'} \int \hat{Q} \int \hat{Q} \int \hat{Q} + ~...~~,
\end{equation}

\noindent
where the integral sign represents a generalized folding operation, 
and $\hat{Q'}$ is obtained from $\hat{Q}$ by removing terms at the
first order in the $NN$ potential \cite{Kuo71,Kuo90}.
The folded-diagram series is then summed up to all orders using the
Lee-Suzuki iteration method \cite{Suzuki80}.

In Ref. \cite{Coraggio12a} the values of the SP energies and TBME
derived including only the N$^3$LO two-body force have been reported.

As shown in the above paper, the diagonalization of $H_{\rm eff}$
performed for some $p$-shell nuclei, such as $^{6}$Li and $^{10}$B,
provides excitation spectra that are in a close agreement with those
obtained in NCSM calculations starting from the same N$^3$LO two-body
potential \cite{Navratil07a,Maris13}.

\begin{figure}[h]
\begin{center}
\includegraphics[scale=0.60,angle=0]{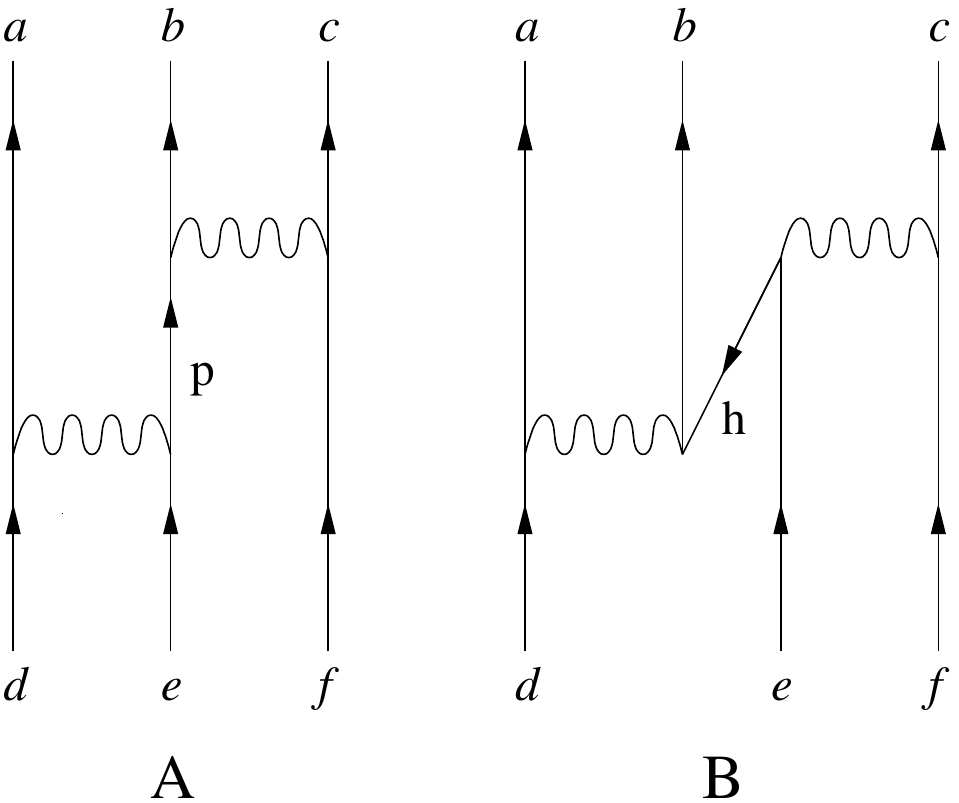}
\caption{Second-order three-body diagrams. The sum over the
  intermediate lines runs over particle and hole states outside the
  model space, shown by A and B, respectively. For the sake of
  simplicity, for each topology we report only one of the diagrams
  which correspond to the permutations of the external lines.}
\label{diagram3corr}
\end{center}
\end{figure}

As regards the calculated g.s. energies, with respect to the $^4$He
core, the agreement between RSM and NCSM deteriorates when increasing
the number of valence nucleons in the shell-model calculations.
This may be ascribed to the fact that our SM Hamiltonian is derived
just for one- and two-valence nucleon systems, and, as mentioned in
the Introduction, for nuclei with more valence nucleons, the
$\hat{Q}$-box should contain diagrams with at least three incoming and
outcoming valence particles.
The leading terms of such correlation diagrams appear at second order
in perturbation theory for three-valence nucleon systems, and are
reported in Fig. \ref{diagram3corr}.

The explicit expressions of these three-body diagrams are reported in
Ref. \cite{Polls83}.
Since the inclusion of a three-body term in the shell-model
Hamiltonian cannot be managed by the SM code we employ \cite{ANTOINE},
we calculate the contribution of the monopole component of the
three-body diagrams and add it to the calculated g.s. energies.

As already mentioned in the Introduction, we calculate $H_{\rm
  eff}$ introducing also the contributions of a N$^2$LO $3N$
potential.
More precisely, we evaluate its contribution at first-order in
many-body perturbation theory only for the one- and two-valence
nucleon systems.

\begin{figure}[h]
\begin{center}
\includegraphics[scale=1.0,angle=0]{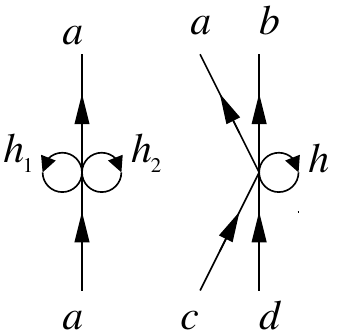}
\caption{First-order one- and two-body diagrams with a
  three-body-force vertex. See text for details}
\label{1b2b3bf}
\end{center}
\end{figure}

As regards the contribution to the single-particle component of
$\hat{Q}$-box from a three-body potential we report in
Fig. \ref{1b2b3bf} the diagram at first order, whose explicit
expression is:

\begin{align}
\label{1b3bfeq}
&\langle j_a| 1{\rm b}_{3N} | j_a \rangle =  ~\nonumber \\
&\mathlarger{\sum}_{\substack{h_1,h_2{}\\J_{12}J}} ~
\frac{\hat{J}^2}{2 \hat{j_a}^2} \langle \left[
  (j_{h_1} j_{h_2})_{J_{12}},j_a \right]_{J}| V_{3N} | \left[
  (j_{h_1} j_{h_2})_{J_{12}},j_a \right]_{J} \rangle~~.
\end{align}

\noindent
The expression of the first-order two-body diagram with a $3N$ vertex,
shown in Fig. \ref{1b2b3bf}, is the following:

\begin{align}
\label{2b3bfeq}
&\langle (j_a j_b)_J| 2{\rm b}_{3N} | (j_c j_d)_J \rangle =  ~\nonumber \\
&\mathlarger{\sum}_{h,J'} ~
\frac{\hat{J'}^2}{\hat{J}^2} \langle \left[
  (j_{a} j_{b})_{J},j_h \right]_{J'}| V_{3N} | \left[
  (j_{c} j_{d})_{J},j_h \right]_{J'} \rangle~~,
\end{align}

The three-body matrix element (3BME) $\langle \left[ (j_{a}
  j_{b})_{J_{ab}},j_c \right]_{J}| V_{3N} | \left[ (j_{d}
  j_{e})_{J_{de}},j_f \right]_{J} \rangle$, expressed within the
proton-neutron formalism, is antisymmetrized but not normalized.

It is worth mentioning that the expressions in Eqs. (\ref{1b3bfeq}) and
(\ref{2b3bfeq}) are the coefficients which multiply the one-body and
two-body terms, respectively, arising from the normal-ordering
decomposition of the three-body component of a many-body
Hamiltonian \cite{HjorthJensen17}.

In the Supplemental Material \cite{supplemental2018}, the calculated
SP energies and TBME of our SM Hamiltonians for $A=6,8,10$, and 12 can
be found. 
As pointed out in Ref. \cite{Coraggio12a}, the $A$ dependence of our
$H_{\rm eff}$s, due to Eq. (\ref{smham_int}), affects mostly the
calculated g.s. energies and very weakly the excited spectra.

\section{Results}
\label{results}
In the following subsections, the results of our shell-model
calculations are presented, first those obtained starting from the
N$^3$LO $NN$ potential only, and then the ones including also the
contributions of the N$^2$LO $3N$ potential.
The calculated spectra and binding energies are compared with those
reported in Refs. \cite{Navratil07a,Maris13}, in order to benchmark
our approach against NCSM, and with the corresponding experimental
data.

\begin{figure}[t]
\begin{center}
\includegraphics[scale=0.32,angle=0]{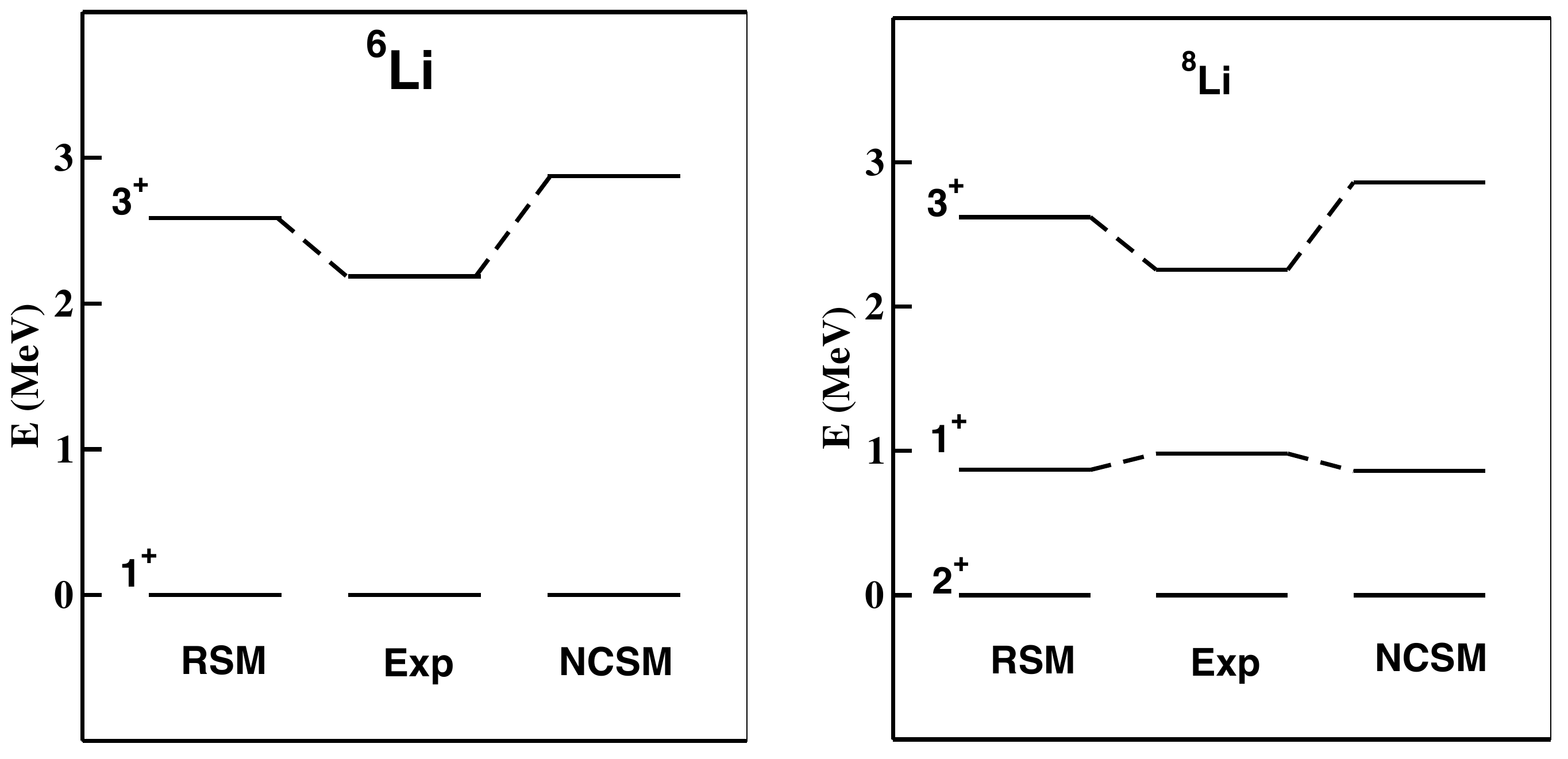}
\caption{Low-lying energy spectra of $^6$Li and $^8$Li. In the middle
  the experimental levels are given, and the calculated ones
  (starting from a two-body potential only) with RSM and NCSM are
  reported on the left and the right side of the figure, respectively.}
\label{SM_NCSM_2b_1}
\end{center}
\end{figure}

\begin{figure}[t]
\begin{center}
\includegraphics[scale=0.32,angle=0]{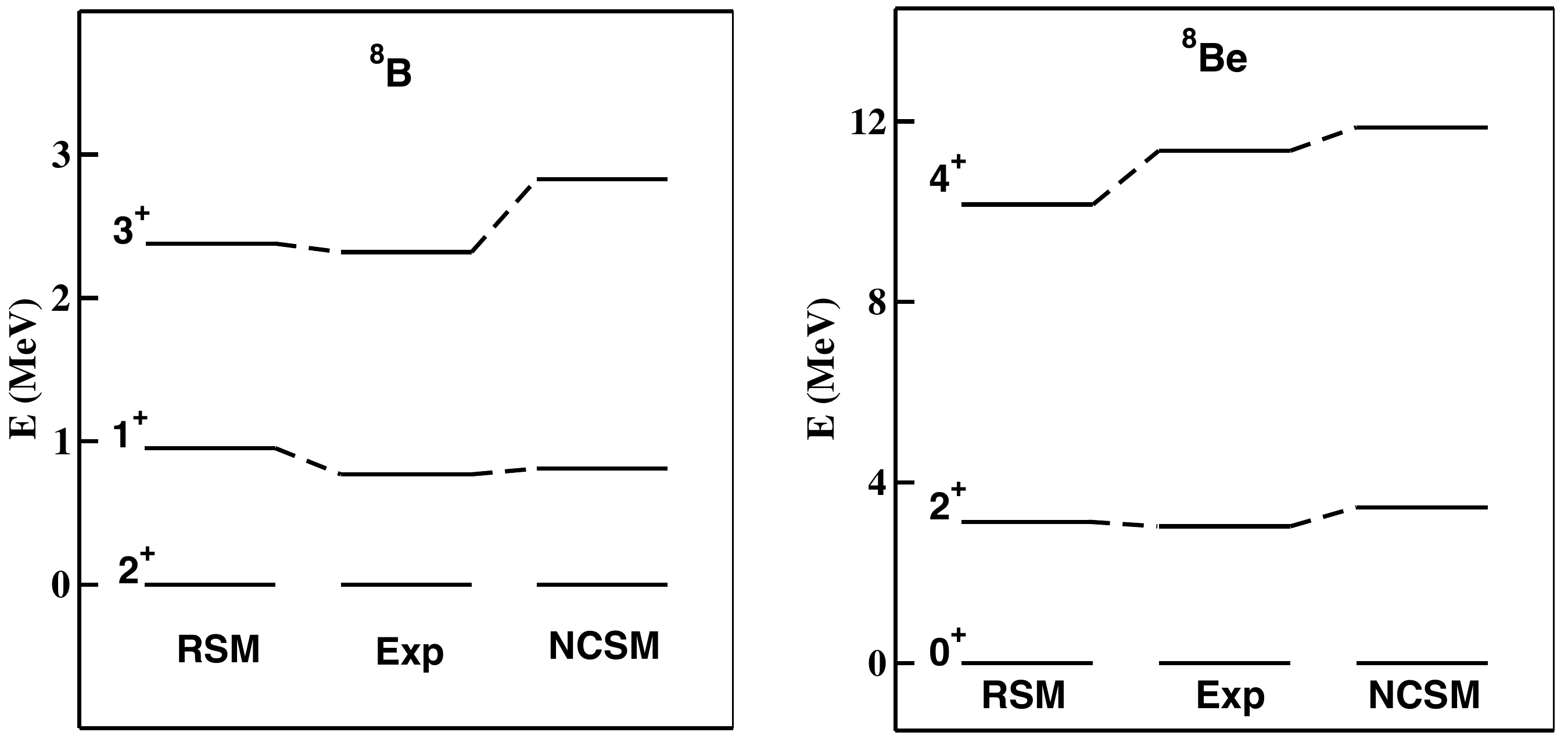}
\caption{Same as Fig. \ref{SM_NCSM_2b_1}, for $^8$B and $^8$Be}
\label{SM_NCSM_2b_2}
\end{center}
\end{figure}

\subsection{Calculations with $NN$ potential}
\label{nnresults}

\begin{figure}[h]
\begin{center}
\includegraphics[scale=0.32,angle=0]{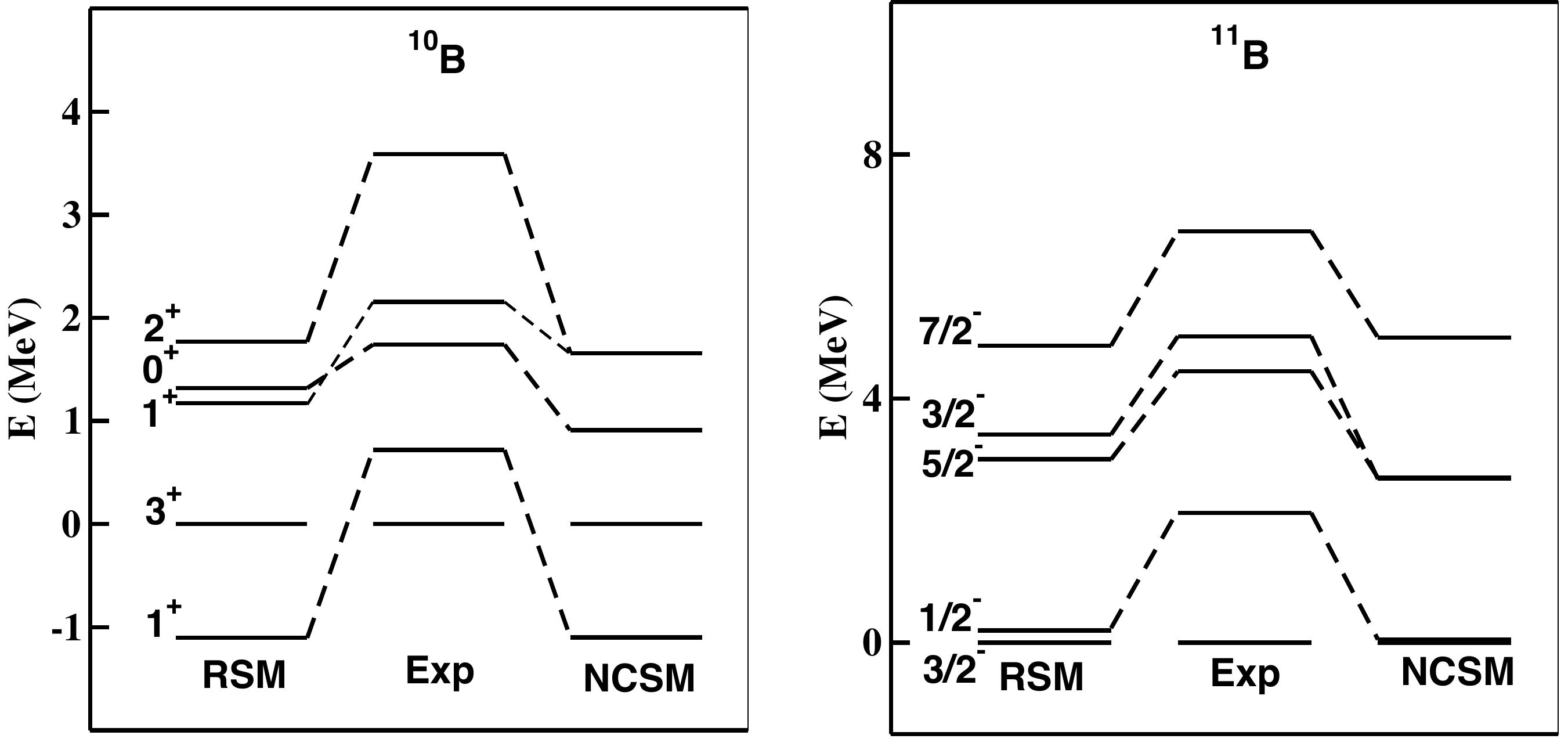}
\caption{Same as Fig. \ref{SM_NCSM_2b_1}, for $^{10}$B and $^{11}$B}
\label{SM_NCSM_2b_3}
\end{center}
\end{figure}

\begin{figure}[h]
\begin{center}
\includegraphics[scale=0.32,angle=0]{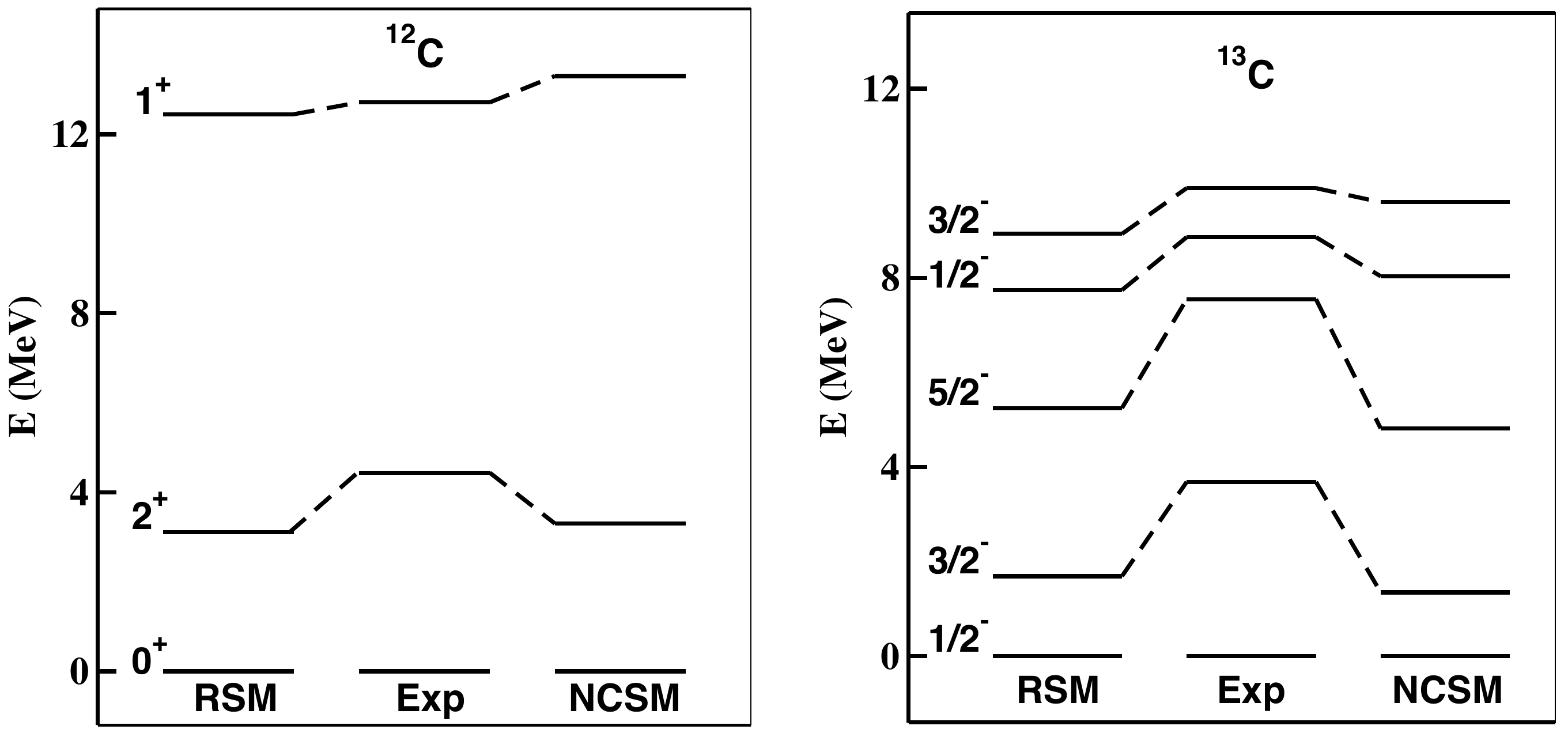}
\caption{Same as Fig. \ref{SM_NCSM_2b_1}, for $^{12}$C and $^{13}$C}
\label{SM_NCSM_2b_4}
\end{center}
\end{figure}

In Figs. \ref{SM_NCSM_2b_1}-\ref{SM_NCSM_2b_4}, the
low-energy spectra of $^6$Li, $^8$Li, $^8$B, $^8$Be, $^{10}$B,
$^{11}$B, $^{12}$C, and $^{13}$C, calculated in our RSM framework are
compared with the experimental ones \cite{ensdf} and those obtained
with NCSM \cite{Navratil07a,Maris13}.
From the inspection of Figs. \ref{SM_NCSM_2b_1}-\ref{SM_NCSM_2b_4}, it
can be seen that there is an excellent agreement between RSM and NCSM,
especially for low-energy levels.

In Fig. \ref{SM_NCSM_2b_3}, we see that both RSM and NCSM predict the
inversion of the $J^\pi =3^+$ g.s. and the first excited $J^\pi=1^+$
state in $^{10}$B.
This defect is healed, as we will see in the next subsection, by
including the $3N$-potential contributions.

As regards $^{11}$B, both RSM and NCSM calculations provide two
low-lying doublets; the almost-degenerate
$J^{\pi}=(\frac{1}{2}^-)_1,(\frac{3}{2}^-)_1$ and
$J^{\pi}=(\frac{3}{2}^-)_2,(\frac{5}{2}^-)_1$ states.
 There is no experimental counterpart of these degeneracies, that will
 be removed including the contribution of a $3N$ potential.

In Fig. \ref{SM_NCSM_2b_4} we report the first-excited states
of $^{12,13}$C isotopes.
It can be seen that both RSM and NCSM fail to reproduce the observed
excitation energy of the yrast $J^{\pi}=2^+$ state in $^{12}$C, that
is underestimated by $\sim$1 MeV.

In Fig. 15 of Ref. \cite{Coraggio12a} we compared our calculated
g.s. energies with respect to $^4$He of $N=Z$ nuclei - up to $^{12}$C
- with those of NCSM calculations \cite{Navratil07a,Maris13}.
Our calculated energies were increasingly underbinding, with respect to the
NCSM ones, and in Ref. \cite{Coraggio12a} we have abscribed this
defect to the lack of many-body components of our $H_{\rm eff}$, whose
role should grow with the number of valence nucleons.

\begin{figure}[h]
\begin{center}
\includegraphics[scale=0.40,angle=0]{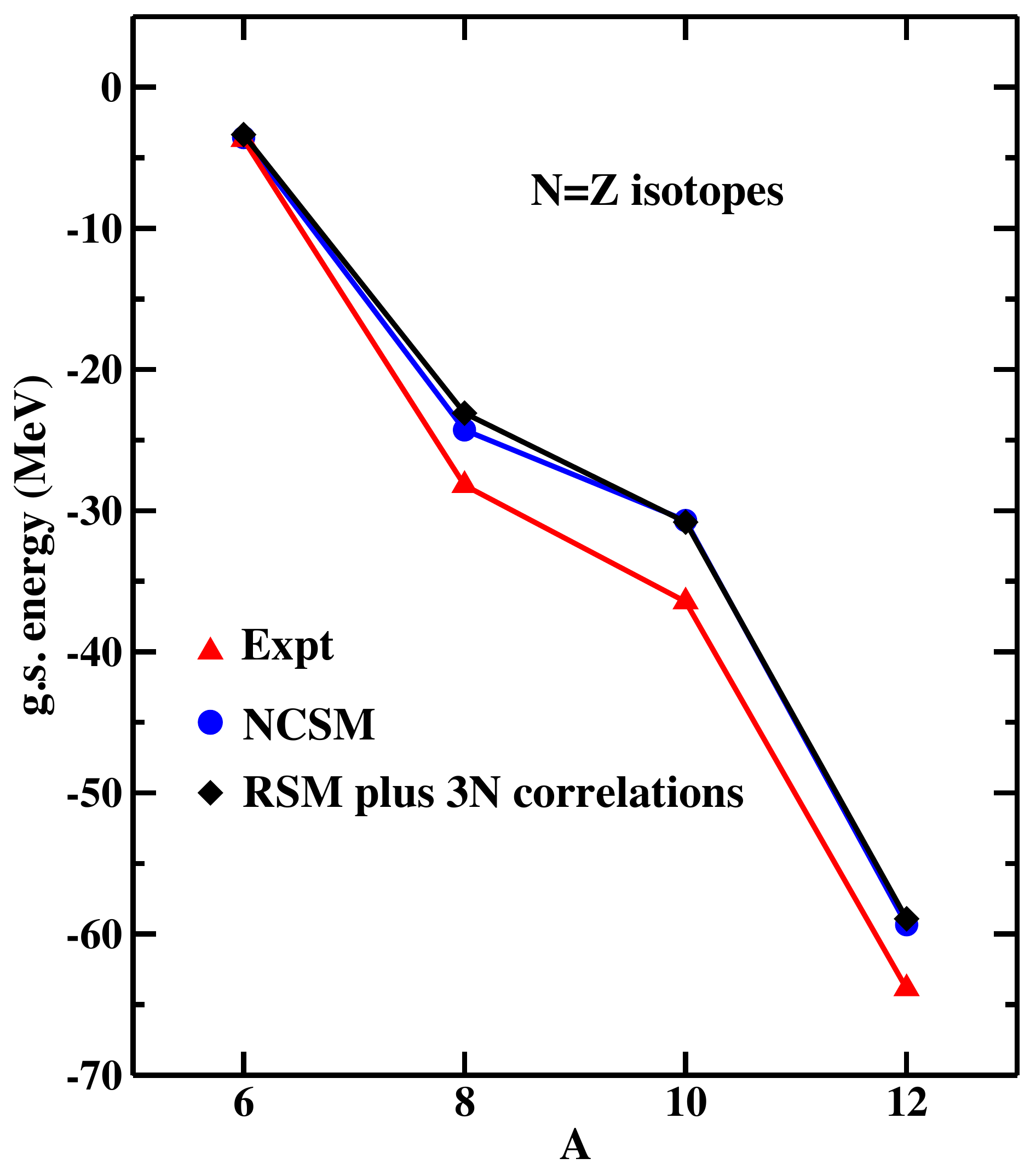}
\caption{(Color online) Ground-state energies for $N=Z$ nuclei with mass $6 \leq A
  \leq 12$.}
\label{gse_2bf}
\end{center}
\end{figure}

As we have mentioned in Section \ref{outline}, we can now include the
three-body diagrams in Fig. \ref{diagram3corr} by calculating their
monopole components and then adding their contributions to the
calculated g.s. energies.
The results of this procedure are reported in Fig. \ref{gse_2bf},
where the new calculated RSM g.s. energies (black squares) are compared with
both the experimental ones (red triangles) and those obtained with
NCSM (blue bullets).
As it can be seen, we have efficiently improved the comparison between
RSM and NCSM, the largest discrepancy being about $4\%$ for
$^8$Be.

\subsection{Calculations with $NN$ plus $3N$ potentials}
\label{nnnresults}

\begin{figure}[h]
\begin{center}
\includegraphics[scale=0.32,angle=0]{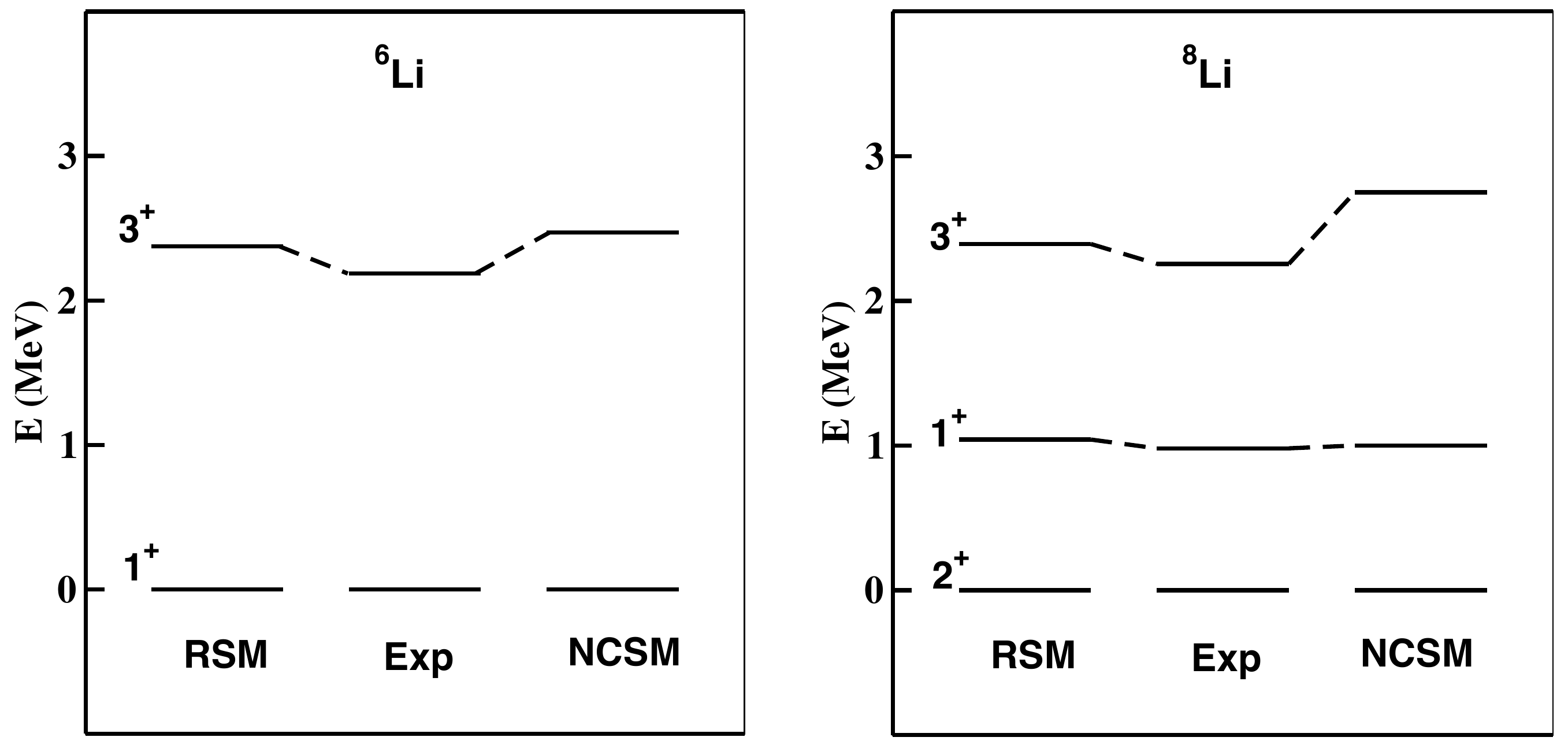}
\caption{Same as Fig. \ref{SM_NCSM_2b_1} but both RSM and NCSM include the
  N$^2$LO $3N$ potential.}
\label{SM_NCSM_3b_1}
\end{center}
\end{figure}

\begin{figure}[h]
\begin{center}
\includegraphics[scale=0.32,angle=0]{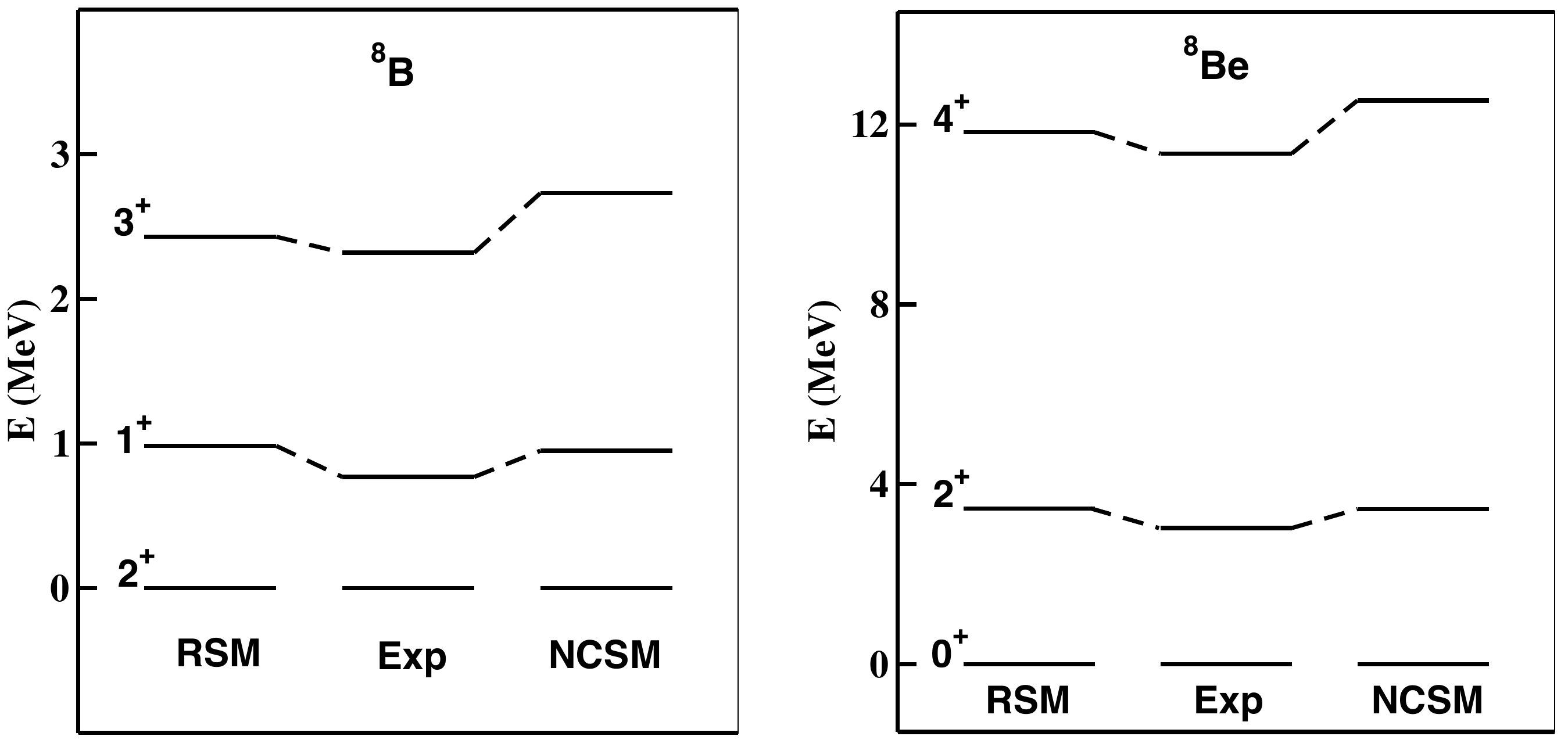}
\caption{Same as Fig. \ref{SM_NCSM_3b_1}, for $^8$B and $^8$Be}
\label{SM_NCSM_3b_2}
\end{center}
\end{figure}

\begin{figure}[h]
\begin{center}
\includegraphics[scale=0.32,angle=0]{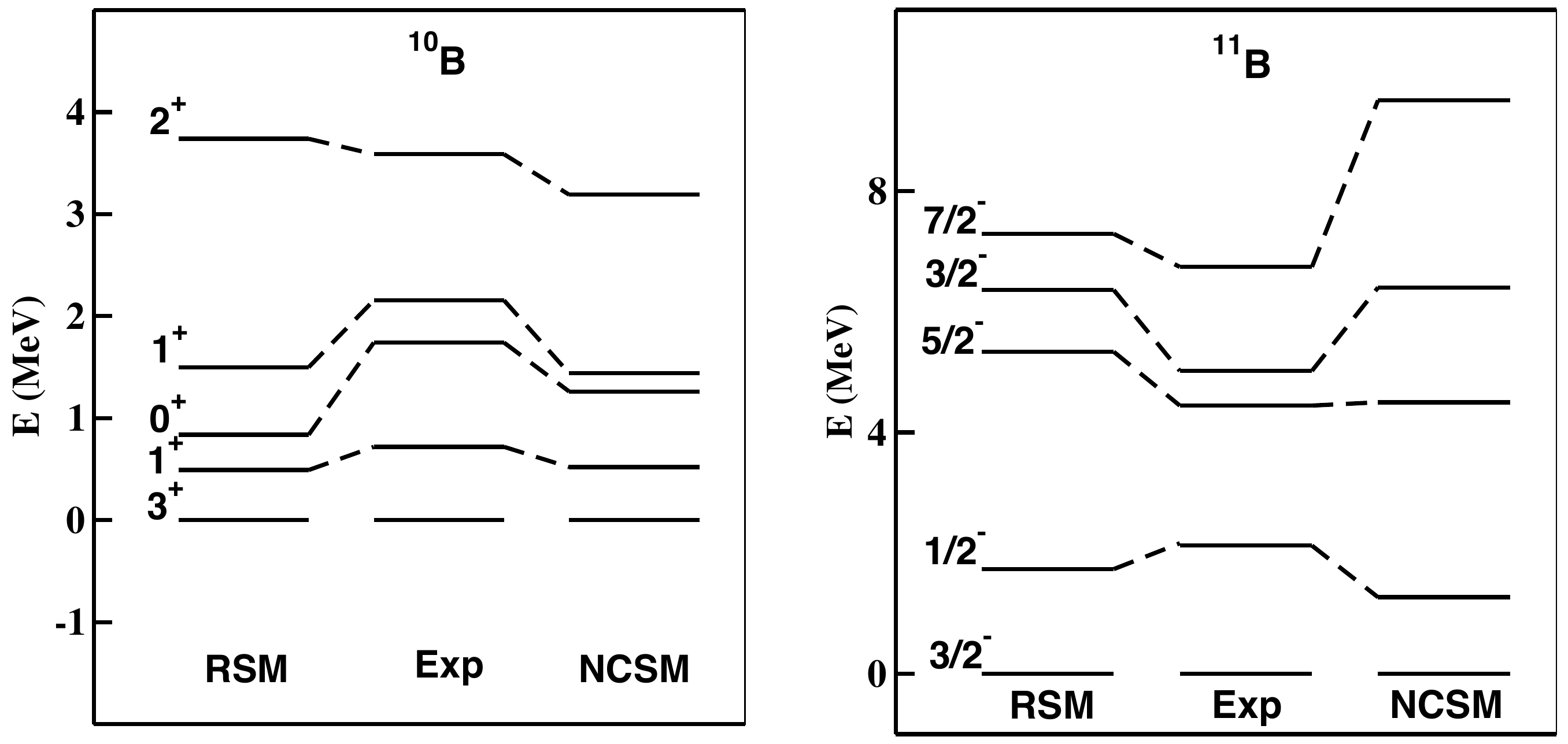}
\caption{Same as Fig. \ref{SM_NCSM_3b_1}, for $^{10}$B and $^{11}$B}
\label{SM_NCSM_3b_3}
\end{center}
\end{figure}

\begin{figure}[h]
\begin{center}
\includegraphics[scale=0.32,angle=0]{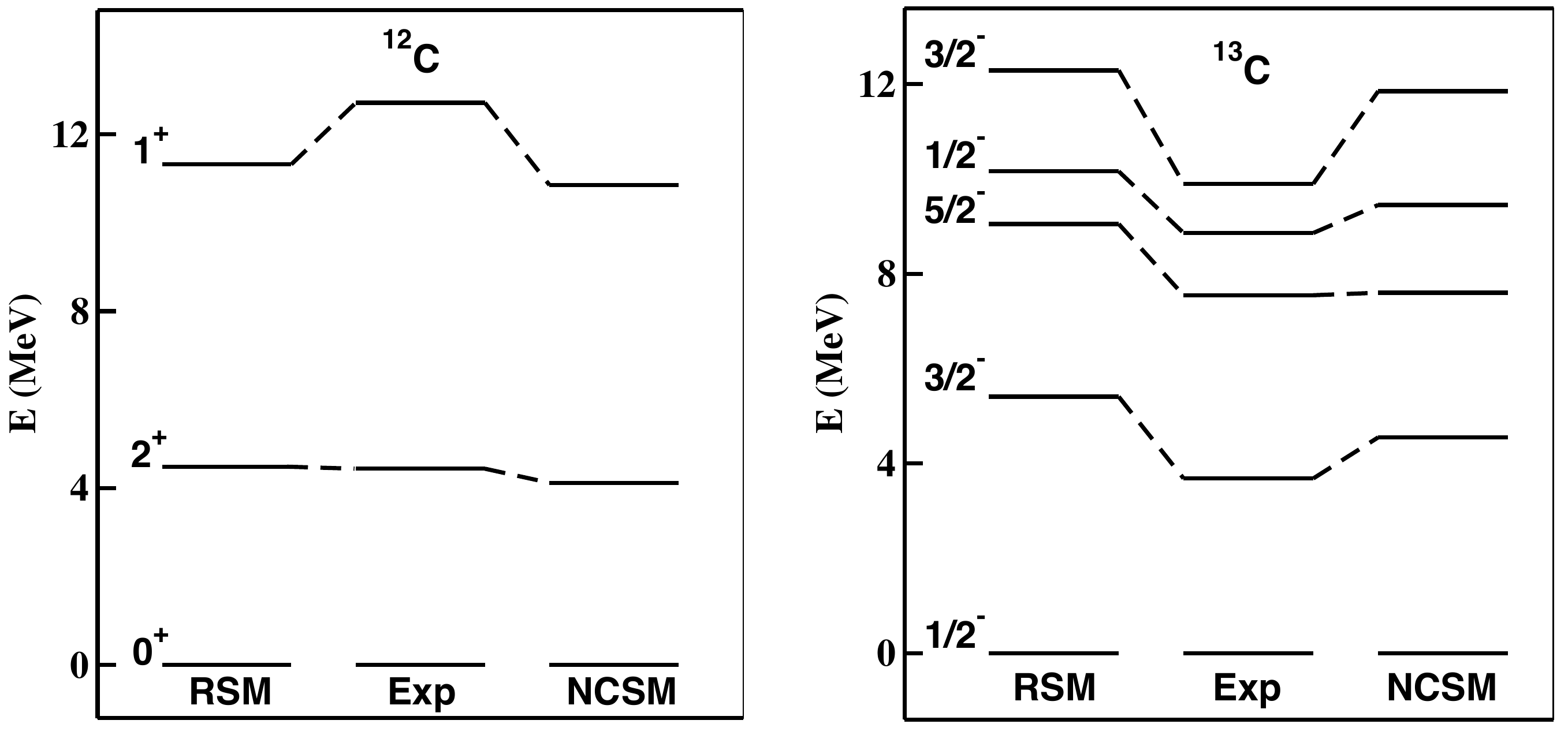}
\caption{Same as Fig. \ref{SM_NCSM_3b_1}, for $^{12}$C and $^{13}$C.}
\label{SM_NCSM_3b_4}
\end{center}
\end{figure}

In Figs. \ref{SM_NCSM_3b_1}-\ref{SM_NCSM_3b_4}, we show the
low-energy spectra of $^6$Li, $^8$Li, $^8$B, $^8$Be, $^{10}$B,
$^{11}$B, $^{12}$C, and $^{13}$C, calculated in our RSM framework, now
including also the contributions from the N$^2$LO $3N$ potential as
reported in Section \ref{outline}.
We compare them with the experimental ones \cite{ensdf} and the NCSM
results \cite{Navratil07a,Maris13}.

As in the case with only N$^3$LO $NN$ potential, our results and NCSM
ones are in a close agreement.
Moreover, the theory with $3N$ compares far better with experiment, as
can be seen in all the reported spectra.

In particular, the experimental sequence of observed states in
$^{10}$B is restored, and the degeneracies of
$J^{\pi}=(\frac{1}{2}^-)_1,(\frac{3}{2}^-)_1$ and
$J^{\pi}=(\frac{3}{2}^-)_2,(\frac{5}{2}^-)_1$ states in $^{11}$B is
removed.
This supports the crucial role played by the $3N$ potential to
improve the spectroscopic description of $p$-shell nuclei.

We recall here that the ESPE are related to the monopole part of the
shell-model hamiltonian, thus reflecting the angular-momentum-averaged
effects of the shell-model interaction $V^{\rm SM}$ for a given nucleus. 
The ESPE of a level is defined as the one-neutron separation
energy of this level, and is calculated in terms of the bare
$\epsilon_j$ and the monopole part of the interaction, namely ${\rm
  ESPE}(j) = \epsilon_j + \sum_{j'} V^{\rm SM}_{j j'} n_{j'}$, where
the sum runs on the model-space levels $j'$, $n_{j}$ being the number
of particles in the level $j$ and $V^{\rm SM}_{j j'}$ the
angular-momentum-averaged interaction $V^{\rm SM}_{j j'} = \sum_J (2J
+1) \langle j j' | V ^{\rm SM}| j j' \rangle_J / \sum_J (2J +1)$.

\begin{figure}[h]
\begin{center}
\includegraphics[scale=0.35,angle=0]{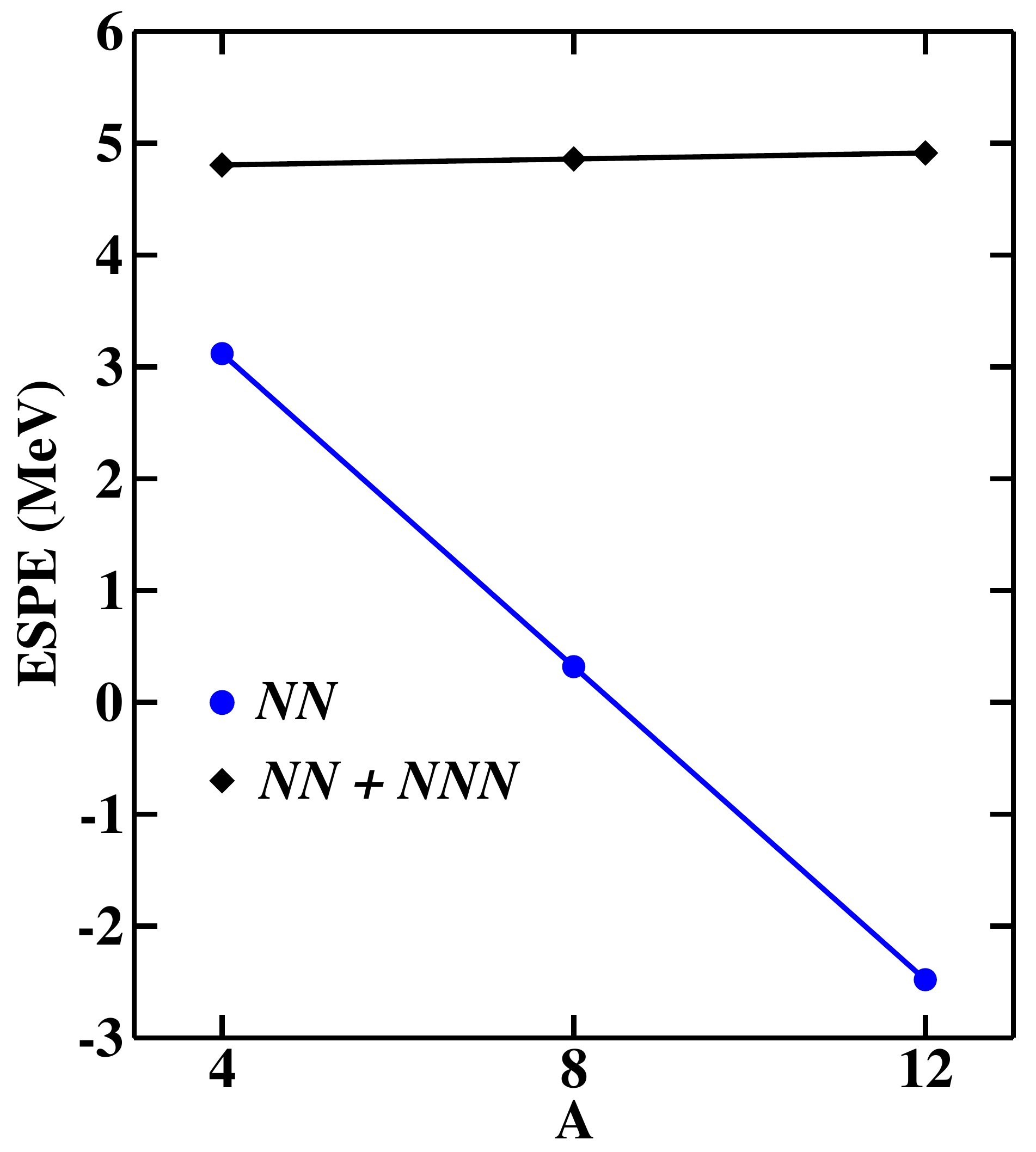}
\caption{(Color online) Proton $0p_{1/2}$ ESPE relative to $0p_{3/2}$ as a
function of $A$ (see text for details). The diamond and bullet symbols refer
to results obtained with and without $3N$ contributions, respectively.}
\label{espep}
\end{center}
\end{figure}

\begin{figure}[h]
\begin{center}
\includegraphics[scale=0.35,angle=0]{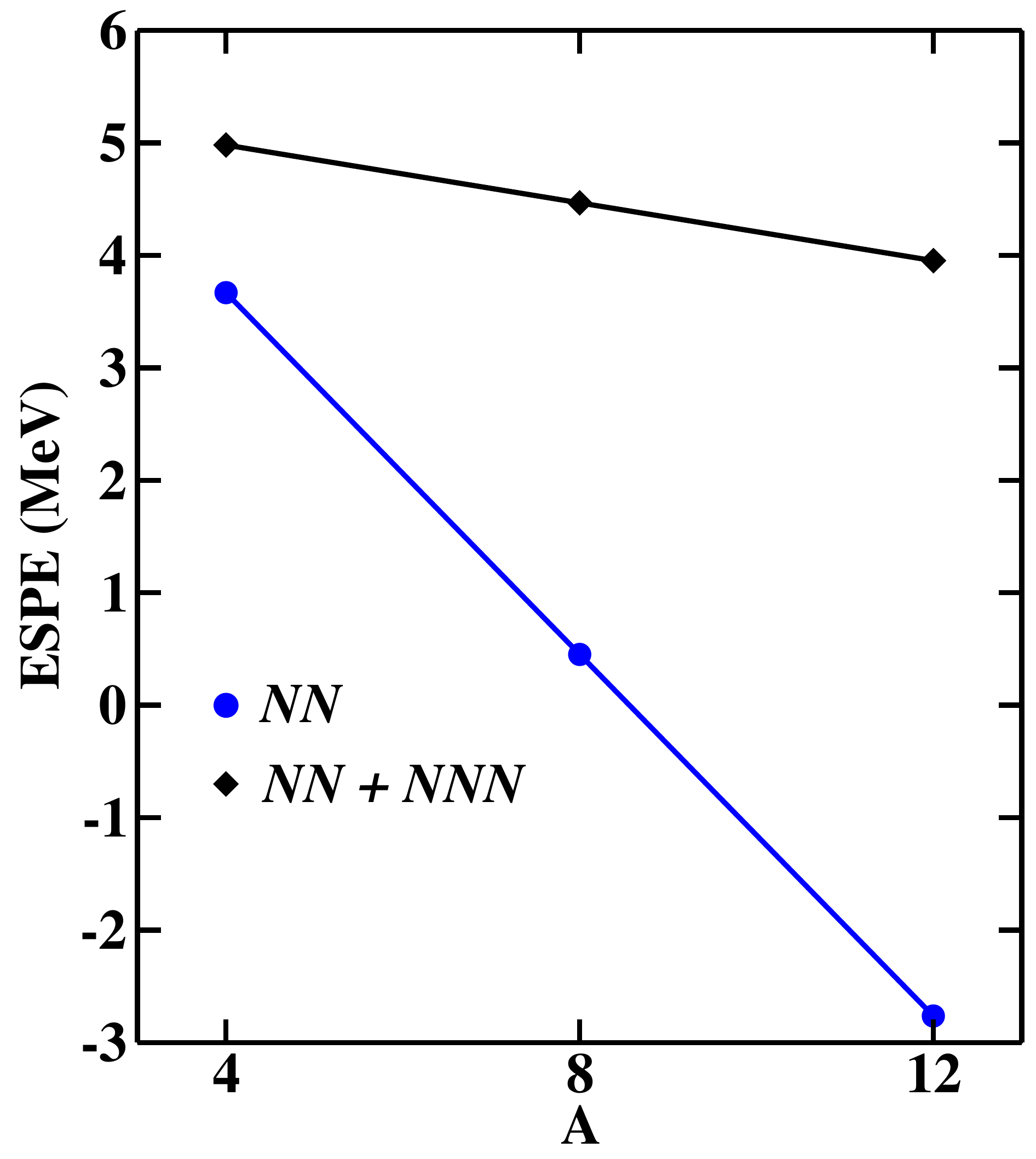}
\caption{(Color online) Same as in Fig. \ref{espep}, but for neutron ESPE.}
\label{espen}
\end{center}
\end{figure}

In Figs. \ref{espep} and \ref{espen}, we show the evolution of proton and
neutron $0p_{1/2}$ ESPE relative to $0p_{3/2}$, respectively, as a
function of $A$ for $N=Z$ isotopes.

The behavior of proton and neutron ESPE is helpful to understand the
different properties of $H_{\rm eff}$, including or not
contributions of the N$^2$LO $3N$ potential.
As can be seen in Figs. \ref{espep} and \ref{espen}, the relative ESPE
rapidly drops down when considering only the N$^3$LO $NN$ potential,
even becoming negative around $A=8$.
Actually, the relative ESPE is almost constant when the $3N$ potential
is taken into account, being $4 \sim 5$ MeV.

\noindent
This reflects in the calculated energy spacings between yrast
$J^{\pi}=\frac{1}{2}^-$ and $\frac{3}{2}^-$ in $^{11}$B and, more
important in the higher excitation energy of the yrast $J^{\pi}=2^+$
state in $^{12}$C spectra reported in Figs. \ref{SM_NCSM_2b_4} and
\ref{SM_NCSM_3b_4}.
In fact, calculations including the three-body force lead
to a better comparison with experiment.

\begin{figure}[ht]
\begin{center}
\includegraphics[scale=0.40,angle=0]{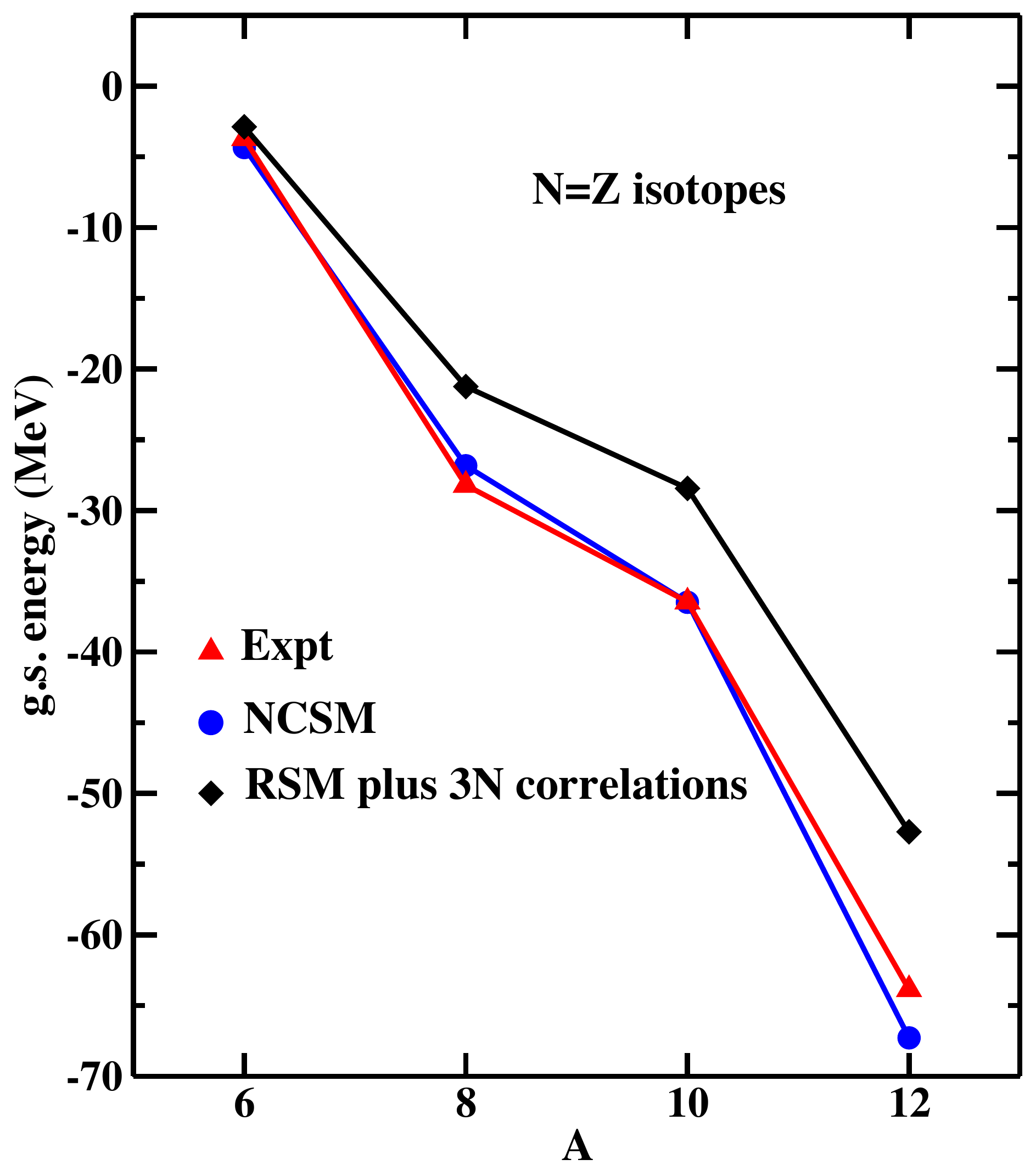}
\caption{(Color online) Same as Fig. \ref{gse_2bf}, but including also
  contributions of N$^2$LO $3N$ potential both in RSM and NCSM
  calculations.}
\label{gse_3bf}
\end{center}
\end{figure}

It is worth pointing out that many studies have been performed about
the crucial role played by three-body potentials on the monopole
properties of SM effective interactions (see for instance
\cite{Zuker03,Schwenk06}), especially to obtain the correct
shell-closure properties when starting from realistic forces.
Recent progress in SM calculations, including three-body
force effects \cite{Otsuka10,Tsunoda17}, has proven the validity of
such a speculation, and our results support the prospect that a
consistent derivation of $H_{\rm eff}$ from chiral two- and three-body
potentials may lead to an improvement of the theoretical description of
nuclear systems with larger mass number $A$. 

For the sake of completeness, in Fig. \ref{gse_3bf} we report the
g.s. energies of $N=Z$ isotopes, that, as in Fig. \ref{gse_2bf},
have been obtained by taking into account also $3N$ correlations. 
Namely, we add to the energies calculated with $H_{\rm eff}$, now
including a N$^2$LO $3N$ potential, only the contribution of their
monopole component. 
It is worth it to point out that the results with RSM are not satisfactory.
Actually, there is a substantial discrepancy between results obtained
with RSM and those calculated with NCSM, since our RSM calculations
underestimate the contribution of the $3N$ potential.
This may be mainly ascribed to the fact that, at present, our $H_{\rm eff}$
includes first-order contributions only.

\section{Concluding remarks and outlook}
\label{conclusions}
In this paper we have presented the results we have obtained for
$p$-shell nuclei in the framework of RSM, taking into account the
contributions of both induced and genuine three-body forces.

On one side, we have shown how the inclusion of the three-body
correlations between valence nucleons that are induced by the $NN$
potential due to the truncation of the Hilbert space greatly improves
the agreement of our calculated binding energies with respect to those
obtained by {\it ab initio} calculations.

On the other side, we have calculated the contribution at first order
in perturbation theory of a N$^2$LO  chiral $3N$ potential to the
SM effective Hamiltonian, and the comparison of our
calculated energy spectra with those from NCSM \cite{Navratil07a,Maris13}
turns out to be successful, thus supporting the reliability of RSM calculations.

Actually, as reported in Sec. \ref{nnnresults}, our g.s. energies -
calculated including the N$^2$LO $3N$ potential - reflect the lack of
higher-order contributions to the perturbative expansion of $H_{\rm eff}$.
This is evidenced by the underestimation of the binding energies
obtained with NCSM, the latter reproducing satisfactorily the observed
ones.

In this regard, the next step of our study will be to include higher-order
contributions with $3N$ vertices in the perturbative expansion of the
$\hat{Q}$ box, in order to establish their role in the evolution of
the spectroscopic properties provided by RSM.

\section*{Acknowledgements}
This work has been supported by the Natural Science Foundation of
China under Grants No. 11320101004, and No. 11575007; and the CUSTIPEN
(China-U.S. Theory Institute for Physics with Exotic Nuclei) funded by
the U.S. Department of Energy, Office of Science under Grant
No. DE-SC0009971.

The authors thank P. Navr\'atil, G. De Gregorio, and T. Miyagi for
helpful comments and fruitful discussions. The calculations have been
carried out at MARCONI of CINECA, Italy.

\appendix*
\section{Calculation of three-body matrix elements}
Our aim is to calculate the three-body matrix elements of the
chiral potential at N$^2$LO between the antisymmetrized three-particle
HO states.
To this end, we follow the procedure sketched below:
\begin{itemize}
\item[(i)] Transformation of the $JT$-coupled three-particle HO basis
  into the Jacobi-HO basis, thus separating the CM and relative motions;
\item[(ii)] antisymmetrization of the Jacobi-HO basis;
\item[(iii)] evaluation of the Jacobi-HO matrix element (ME), which is the 3BME
	     of the chiral interaction at N$^2$LO \cite{Bogner05,Machleidt11}.
\end{itemize}
Our approach to the steps (i) and (ii) is essentially same as that used in Ref.~\cite{Navratil07b}.

The transformation in the step (i) leads to the so-called $T$
coefficients (see, for instance, Ref.~\cite{Roth14}), involving
angular momentum recouplings and the HO brackets
originating from the Talmi transformation~\cite{Talmi52,Brody60,Moshinsky60,Trlifaj72,Buck96}.
The HO brackets are computed by using the Fortran code of
Ref.~\cite{Kamuntavicius01}.

At the step (ii), as suggested in Refs.~\cite{Navratil99,Navratil00c},
we build up the antisymmetrized Jacobi-HO states $\ket{\kappa;JT}_A$ by diagonalizing the three-body antisymmetrizer.
Thus we obtain
\begin{align}
 \ket{\kappa;JT}_A
 &=
 \sqrt{6}\sum_{\bar\kappa}D_{\kappa\bar\kappa}^{(JT)}
 \ket{\bar\kappa;JT},
 \label{Asym}\\
 D_{\kappa\bar\kappa}^{(JT)}
 &=
 \sum_{\eta}\Braket{\kappa;JT| \eta}\Braket{\eta| \bar\kappa;JT},
 \label{CFP}
\end{align}
where $\Ket{\eta}$ is a ``physical'' eigenstate~\cite{Navratil00c} of the three-body antisymmetrizer,
and it corresponds to the eigenvalue 1.
The index $\kappa$ for the totally antisymmetrized Jacobi-HO states $\ket{\kappa;JT}_A$
stands for the set of the quantum numbers
$\left\{n_{12},l_{12},S_{12},I_{12},T_{12},n,l,I\right\}$.
The quantum numbers with the subscript ``12'' are associated with the
$a$-$b$ system, that is, the principal quantum number $n_{12}$, the
orbital angular momentum $l_{12}$, the two-nucleon coupled spin
$S_{12}$, the angular momentum $I_{12}$ originating from the coupling
of $l_{12}$ and $S_{12}$, and the two-nucleon coupled isospin $T_{12}$.
Whereas, the $(ab)$-$c$ motion is characterized by the principal
quantum number $n$, the orbital angular momentum $l$, and the angular
momentum $I$ coming from the coupling of $l$ and the nucleon spin $1/2$.
The total angular momentum $J$ (isospin $T$) is formed by the coupling
of $I_{12}$ and $I$ ($T_{12}$ and nucleon isospin $1/2$).
The index $\bar\kappa=\left\{\bar n_{12},\bar l_{12},\bar S_{12},\bar I_{12},\bar T_{12},\bar n,\bar l,\bar I\right\}$
is similar to $\kappa$ but for the Jacobi-HO states $\ket{\bar\kappa;JT}$,
which is partially antisymmetrized with respect to the $a$-$b$ system
with the constraint $(-1)^{\bar l_{12}+\bar S_{12}+\bar T_{12}}=-1$.

As regards the step (iii), the Jacobi-HO MEs both of the one-pion-exchange plus-contact operator $V_{3N}^{(1\pi)}$
and the $3N$ contact operator $V_{3N}^{(\rm ct)}$ are evaluated with a nonlocal regulator (see Eq.~\eqref{Reg1})
following the procedure described in Ref.~\cite{Navratil07b}.
At variance with Ref.~\cite{Navratil07b}, the Jacobi-HO ME of the two-pion exchange operator $V_{3N}^{(2\pi)}$
is calculated in an alternative way explained below.

Owing to Eq.~\eqref{Asym} and the symmetry of $V_{3N}^{(2\pi)}$ with respect to the permutation of particles,
the antisymmetrized Jacobi-HO ME is given by
\begin{align}
 &{}_{\substack{\\\\A}}\!\Braket{\kappa';JT \left| V_{3N}^{(2\pi)}\right|\kappa;JT}_A
 \nonumber\\
 &\quad=
 18\sum_{\bar\kappa\bar\kappa'}D_{\kappa\bar\kappa}^{(JT)}D_{\kappa'\bar\kappa'}^{(JT)}
 \nonumber\\
 &\quad\times
 \Braket{\bar\kappa';JT \left| \left[W_{3N}^{(2\pi;c_1)}+W_{3N}^{(2\pi;c_3)}+W_{3N}^{(2\pi;c_4)}\right]\right|\bar\kappa;JT},
 \label{2piSym}
\end{align}
The momentum representation of the reduced operator $W_{3N}^{(2\pi;c_\mu)}$, with $\mu=1,~3,~\textrm{or}~4$, is written as
\begin{align}
 \Braket{\!\vect{p}_a'\vect{p}_b'\vect{p}_c'\! \left| W_{3N}^{(2\pi;c_\mu)} \right|\!\vect{p}_a\vect{p}_b\vect{p}_c\!}
 \!=\!w_{3N}^{(2\pi;c_\mu)}(\vect{q}_b,\vect{q}_c)
 \delta(\vect{q}_a\!\!+\!\vect{q}_b\!\!+\!\vect{q}_c).
 \label{2piPotW}
\end{align}
Here, for convenience, we define $w_{3N}^{(2\pi;c_\mu)}$ as a function of $\vect{q}_b$ and $\vect{q}_c$,
because these two transferred momenta are simply written in terms of the Jacobi momenta we employ (See Eq.~\eqref{Jacobimom1}).
The explicit form of $w_{3N}^{(2\pi;c_\mu)}$ is expressed as follows (see, for example Refs.~\cite{Machleidt11,Epelbaum06}):
\begin{align}
 &w_{3N}^{(2\pi;c_1)}(\vect{q}_b,\vect{q}_c)
 \nonumber\\
 &\quad=
 -\frac{1}{(2\pi)^6}\frac{g_A^2c_1m_\pi^2}{f_\pi^4}
 \frac{\left(\vect{\sigma}_b\cdot\vect{q}_b\right)\left(\vect{\sigma}_c\cdot\vect{q}_c\right)}
 {\left(q_b^2+m_\pi^2\right)\left(q_c^2+m_\pi^2\right)}
 \vect{\tau}_b\cdot\vect{\tau}_{c},
 \label{w2pic1}\\
 &w_{3N}^{(2\pi;c_3)}(\vect{q}_b,\vect{q}_c)
 \nonumber\\
 &\quad=
 \frac{1}{(2\pi)^6}\frac{g_A^2c_3}{2f_\pi^4}
 \frac{\left(\vect{\sigma}_b\cdot\vect{q}_b\right)\left(\vect{\sigma}_c\cdot\vect{q}_c\right)}
 {\left(q_b^2+m_\pi^2\right)\left(q_c^2+m_\pi^2\right)}
 \left(\vect{q}_b\cdot\vect{q}_c\right)
 \left(\vect{\tau}_b\cdot\vect{\tau}_{c}\right),
 \label{w2pic3}\\
 &w_{3N}^{(2\pi;c_4)}(\vect{q}_b,\vect{q}_c)
 \nonumber\\
 &\quad=
 \frac{1}{(2\pi)^6}\frac{g_A^2c_4}{4f_\pi^4}
 \frac{\left(\vect{\sigma}_b\cdot\vect{q}_b\right)\left(\vect{\sigma}_c\cdot\vect{q}_c\right)}
 {\left(q_b^2+m_\pi^2\right)\left(q_c^2+m_\pi^2\right)}
 \nonumber\\
 &\quad\times
 \left\{\left(\vect{q}_b\times\vect{q}_c\right)\cdot\vect{\sigma}_{a}\right\}
 \left\{\left(\vect{\tau}_b\times\vect{\tau}_c\right)\cdot\vect{\tau}_{a}\right\}.
 \label{w2pic4}
\end{align}
where $\boldsymbol{\sigma}_i$ ($\boldsymbol{\tau}_i$) is the Pauli spin (isospin) matrix of nucleon $i$~($i=a,~b,~\textrm{or}~c$),
and the transferred momentum is $\vect{q}_i~\!=~\!\vect{p}'_i~\!-~\!\vect{p}_i$, with $\vect{p}_i$ and $\vect{p}'_i$
being the initial and final momenta, respectively.
We use the parameters, $g_A=1.29$, $f_\pi=92.4$ MeV, $m_\pi=138.04$ MeV, and $\Lambda_\chi=700$ MeV.
In this paper the parameters are given in natural units, namely $c=\hbar=1$.
Note that, in $w_{3N}^{(2\pi;c_\mu)}$ there is a prefactor $1/(2\pi)^6$, which
originates from our convention of the normalization,
$\Braket{\vect{p}_a'\vect{p}_b'\vect{p}_c' | \vect{p}_a\vect{p}_b\vect{p}_c }
=\delta(\vect{q}_a)\delta(\vect{q}_b)\delta(\vect{q}_c)$.
See Refs.~\cite{Navratil07b,Coon81} for more details.

Although a local regulator depending on $\vect{q}_i$ is adopted in Ref.~\cite{Navratil07b},
alternatively we employ a nonlocal regulator,
\begin{align}
 u_\nu\!\left(k,K,\Lambda_0\right)
 = \exp\left[-\!\left(\frac{k^2+K^2}{2\Lambda_0^2}\right)^{\!\!\nu}\,\right],
 \label{Reg1}
\end{align}
which is consistent with that for the two-body N$^3$LO potential with $\Lambda_0=500$~MeV and $\nu=2$.
The Jacobi momenta $\vect{k}$ and $\vect{K}$ are defined by
\begin{align}
 \vect{k}=\frac{1}{\sqrt{2}}\left(\vect{p}_a\!\!-\!\vect{p}_b\right),\quad
 \vect{K}=\sqrt{\frac{2}{3}}\!\left[\frac{1}{2}\left(\vect{p}_a\!\!+\!\vect{p}_b\right)\!-\!\vect{p}_c\right].
 \label{Jacobimom1}
\end{align}
Thus we regularize $w_{3N}^{(2\pi;c_\mu)}$ as
\begin{align}
 &w_{3N}^{(2\pi;c_\mu)}(\vect{q}_b,\vect{q}_c)
 \nonumber\\
 &\quad\to
 u_\nu\!\left(k',K',\Lambda_0\right)
 w_{3N}^{(2\pi;c_\mu)}(\vect{q}_b,\vect{q}_c)
 u_\nu\!\left(k,K,\Lambda_0\right),
 \label{2piPotReg}
\end{align}
and we express it in terms of $k$, $k'$, $K$, $K'$, $\cos\theta_1$, $\cos\theta_2$, and $\cos\theta_3$,
where the prime stands for the Jacobi momenta in the
final channel and $\theta_1$, $\theta_2$, and $\theta_3$ are the
angles between $\vect{K}$ and $\vect{K}'$, $\vect{k}$ and $\vect{k}'$,
and $\vect{K}-\vect{K}'$ and $\vect{k}-\vect{k}'$, respectively.
Successively, we perform the triple-fold multipole expansion for these angles.
As a result, we obtain the regularized 3BMEs of each operator as
\begin{align}
 &\Braket{\bar\kappa';JT \left| W_{3N}^{(2\pi;c_1)}\right|\bar\kappa;JT}
 \nonumber \\
 &\quad =
 3 c_1 m_\pi^2
 S_{\bar\kappa \bar\kappa'}^{JT}
\begin{Bmatrix}
  \bar S_{12} & \bar S_{12}' & 1 \\
  \frac{1}{2} & \frac{1}{2}  & \frac{1}{2}
\end{Bmatrix}
\begin{Bmatrix}
  \bar T_{12} & \bar T_{12}' & 1 \\
  \frac{1}{2} & \frac{1}{2}  & \frac{1}{2}
\end{Bmatrix}
 \nonumber\\
 &\quad \times
 \sum_{\substack{\lambda_b \lambda_c \\\lambda_b'\lambda_b''}}
 \sum_{\substack{\lambda_1\lambda_2\lambda_3\\\lambda_3'\lambda_3''}} 
 \sum_{l_1} (-1)^{\lambda_b+l_1+1} \hat{l}_1^2  \nonumber\\
&\quad \times I_{\bar\kappa \bar\kappa' L_b=2,L_c=2, L_b'=1,L_c'=1}
   ^{\nu\lambda_b\lambda_c\lambda_b'\lambda_b''\lambda_1\lambda_2\lambda_3\lambda_3'\lambda_3''}\!\left(\Lambda_0\right)
   \nonumber \\
&\quad \times
 X_{\bar\kappa \bar\kappa' J, L_0=1,L_b'=1,L_c'=1,l_0=\lambda_b,l_1}
 ^{\lambda_b\lambda_c\lambda_b'\lambda_b''\lambda_1\lambda_2\lambda_3\lambda_3'\lambda_3''},
\label{ME2pic1}
\end{align}
\begin{align}
 &\Braket{\bar\kappa';JT \left|W_{3N}^{(2\pi;c_3)}\right|\bar\kappa;JT}
 \nonumber\\
 &\quad=
 \frac{\sqrt{3}}{2}c_3
 S_{\bar\kappa \bar\kappa'}^{JT}
 \begin{Bmatrix}
  \bar S_{12} & \bar S_{12}' & 1 \\
  \frac{1}{2} & \frac{1}{2}  & \frac{1}{2}
 \end{Bmatrix}
 \begin{Bmatrix}
  \bar T_{12} & \bar T_{12}' & 1 \\
  \frac{1}{2} & \frac{1}{2}  & \frac{1}{2}
 \end{Bmatrix}
 \nonumber\\
 &\quad\times
 \sum_{L_bL_c}
 \sum_{\substack{\lambda_b \lambda_c \\\lambda_b'\lambda_b''}}
 \sum_{\substack{\lambda_1\lambda_2\lambda_3\\\lambda_3'\lambda_3''}}
 \sum_{l_0l_1}
 \hat{L}_b\hat{L}_c
 \hat{l}_0^2\hat{l}_1^2
 \left(1 0 1 0 | L_b 0\right)
 \nonumber\\
 &\quad \times \left(1 0 1 0 | L_c 0\right)
 \begin{Bmatrix}
  L_b\!-\!\lambda_b & \lambda_b & L_b \\
  1                 & 1         & l_0
 \end{Bmatrix}
 \begin{Bmatrix}
  l_0 & l_1 & 1         \\
  L_c & 1   & \lambda_b
 \end{Bmatrix}
 \nonumber\\
 &\quad\times
 I_{\bar\kappa \bar\kappa' L_bL_c,L_b'=L_b,L_c'=L_c}
  ^{\nu\lambda_b\lambda_c\lambda_b'\lambda_b''\lambda_1\lambda_2\lambda_3\lambda_3'\lambda_3''}\!\left(\Lambda_0\right)
  \nonumber \\
&\quad \times X_{\bar\kappa \bar\kappa' J, L_0=1,L_b'=L_b,L_c'=L_c,l_0l_1}
 ^{\lambda_b\lambda_c\lambda_b'\lambda_b''\lambda_1\lambda_2\lambda_3\lambda_3'\lambda_3''},
 \label{ME2pic3}
\end{align}
\begin{align}
 &\Braket{\bar\kappa';JT \left|W_{3N}^{(2\pi;c_4)}\right|\bar\kappa;JT}
 \nonumber\\
 &\quad=
 9\sqrt{3}c_4
 (-)^{l_{12}'+1}S_{\bar\kappa \bar\kappa'}^{JT}
 \begin{Bmatrix}
  \frac{1}{2} & \frac{1}{2} & \bar T_{12}' \\[3pt]
  \frac{1}{2} & \frac{1}{2} & \bar T_{12}  \\[3pt]
  1           & 1           & 1
 \end{Bmatrix}
 \nonumber\\
 &\quad\times
 \sum_{\substack{L_0\\L_bL_c}}
 \sum_{\substack{\lambda_b \lambda_c \\\lambda_b'\lambda_b''}}
 \sum_{\substack{\lambda_1\lambda_2\lambda_3\\\lambda_3'\lambda_3''}}
 \sum_{l_0l_1}
 \hat{L}_0^2\hat{L}_b\hat{L}_c \hat{l}_0^2\hat{l}_1^2
 \left(1 0 1 0 | L_b 0\right) \nonumber \\
 &\quad \times \left(1 0 1 0 | L_c 0\right)
 \begin{Bmatrix}
  L_0 & L_b & 1 \\
  1   & 1   & 1
 \end{Bmatrix}
 \begin{Bmatrix}
  L_b\!-\!\lambda_b & \lambda_b & L_b \\
  1             & L_0       & l_0
 \end{Bmatrix} \nonumber \\
 &\quad \times
 \begin{Bmatrix}
  l_0 & l_1 & 1 \\
  L_c & 1   & \lambda_b
 \end{Bmatrix}
 \begin{Bmatrix}
  \frac{1}{2} & \frac{1}{2} & \bar S_{12}' \\[3pt]
  \frac{1}{2} & \frac{1}{2} & \bar S_{12}  \\[3pt]
  1           & 1           & L_0
 \end{Bmatrix} \nonumber \\
 &\quad \times
 I_{\bar\kappa \bar\kappa' L_bL_c,L_b'=L_b,L_c'=L_c}
  ^{\nu\lambda_b\lambda_c\lambda_b'\lambda_b''\lambda_1\lambda_2\lambda_3\lambda_3'\lambda_3''}\!\left(\Lambda_0\right)
 \nonumber\\
 &\quad\times
 X_{\bar\kappa \bar\kappa' J, L_0L_b'=L_b,L_c'=L_c,l_0l_1}
 ^{\lambda_b\lambda_c\lambda_b'\lambda_b''\lambda_1\lambda_2\lambda_3\lambda_3'\lambda_3''},
 \label{ME2pic4}
\end{align}
where, in general, $\hat{x}=\sqrt{2x+1}$.
The coefficients in Eqs.~\eqref{ME2pic1}-\eqref{ME2pic4} are defined as
\begin{align}
 S_{\bar\kappa \bar\kappa'}^{JT}
 &=
 \left[\frac{g_A}{\left(\pi f_\pi\right)^2}\right]^2
 i^{\bar l_{12}+\bar l_{12}'+\bar l+\bar l'}
 \nonumber\\
 &\times
 (-)^{\bar S_{12}+\bar I_{12}'-\bar I+\mathcal{I}+\bar T_{12}+\bar T_{12}'+T+\frac{1}{2}}
 \nonumber\\
 &\times
 \hat{\bar S}_{12}\hat{\bar S}_{12}'\hat{\bar I}_{12}\hat{\bar I}_{12}'\hat{\bar I}\hat{\bar I}'\hat{\bar T}_{12}\hat{\bar T}_{12}'
 \begin{Bmatrix}
  \bar T_{12} & \bar T_{12}' & 1 \\
  \frac{1}{2} & \frac{1}{2}  & T
 \end{Bmatrix},
 \label{Scommon}
\end{align}
\begin{align}
 &I_{\bar\kappa \bar\kappa' L_bL_cL_b'L_c'}
   ^{\nu \lambda_b\lambda_c\lambda_b'\lambda_b''\lambda_1\lambda_2\lambda_3\lambda_3'\lambda_3''}\!\left(\Lambda_0\right)
 \nonumber\\
 &\quad=
 3^{-\frac{\lambda_b}{2}}(-1)^{\lambda_b+\lambda_c+\lambda_b'+\lambda_b''+\lambda_1+\lambda_2+\lambda_3+\lambda_3'+\lambda_3''}
 \nonumber\\
 &\quad\times
 \Widehat{L_b'\!\!-\!\!\lambda_b}\,\,\Widehat{L_c'\!\!-\!\!\lambda_c}\,\,\Widehat{L_b'\!\!-\!\!\lambda_b\!\!-\!\!\lambda_b'}\,\,
 \Widehat{\lambda_b\!\!-\!\!\lambda_b''}\,\,
 \Widehat{\lambda_3\!\!-\!\!\lambda_3'}\,\,\Widehat{\lambda_3\!\!-\!\!\lambda_3''}
 \nonumber\\
 &\quad\times
 \left[C^{2L_b'+1}_{2\lambda_b}C^{2L_c'+1}_{2\lambda_c}C^{2(L_b'-\lambda_b)+1}_{2\lambda_b'}C^{2\lambda_b+1}_{2\lambda_b''}
 C^{2\lambda_3+1}_{2\lambda_3'}C^{2\lambda_3+1}_{2\lambda_3''}\right]^{\frac{1}{2}}
 \nonumber\\
 &\quad\times
 \int\!\!\!\!\int\!\!\!\!\int\!\!\!\!\int dk dK dk' dK'
 f_{\lambda_1\lambda_2\lambda_3}^{(L_bL_c)}(k,k',K,K')
 \nonumber\\
 &\quad\times
 k^{L_b'-\lambda_b-\lambda_b'+\lambda_3-\lambda_3'+1}
 K^{L_c'-\lambda_c+\lambda_b-\lambda_b''+\lambda_3-\lambda_3''+1}
 \nonumber\\
 &\quad\times
 k'^{\lambda_b'+\lambda_3'+1}
 K'^{\lambda_c+\lambda_b''+\lambda_3''+1}
 \nonumber\\
 &\quad\times
 P_{\bar n_{12}\bar l_{12}}\!\left(k\right)P_{\bar n\bar l}\!\left(K\right)
 P_{\bar n_{12}'\bar l_{12}'}\!\left(k'\right)P_{\bar n'\bar l'}\!\left(K'\right)
 \nonumber\\
 &\quad\times
 u_\nu\!\left(k,K,\Lambda_0\right)
 u_\nu\!\left(k',K',\Lambda_0\right),
 \label{Icommon}
\end{align}
\begin{align}
 &X_{\bar\kappa \bar\kappa' J L_0L_b'L_c'l_0l_1}
   ^{\lambda_b\lambda_c\lambda_b'\lambda_b''\lambda_1\lambda_2\lambda_3\lambda_3'\lambda_3''}
 \nonumber\\
 &\quad=
 \sum_{l_2l_3}
 \sum_{\substack{\lambda\lambda'\\\Lambda\Lambda'}}
 \sum_{L_1L_2L_3}
 (-1)^{L_1+L_2+L_3}
 \hat{l}_2\hat{l}_3
 \hat{\lambda}\hat{\lambda}'\hat{\Lambda}\hat{\Lambda}'
 \hat{L}_1^2\hat{L}_2^2\hat{L}_3^2
 \nonumber\\
 &\quad\times
 \left(L_c'\!-\!\lambda_c,0,\lambda_b\!-\!\lambda_b'',0|l_2 0\right)
 \left(\lambda_c 0 \lambda_b''0|l_3 0\right)
 \nonumber\\
 &\quad\times
 \left(L_b'\!-\!\lambda_b\!-\!\lambda_b', 0 \lambda 0 | \bar l_{12} 0\right)
 \left(\lambda_b' 0 \lambda' 0 | \bar l_{12}' 0\right)
 \nonumber\\
 &\quad\times
 \left(l_2 0 \Lambda 0 | \bar l 0\right)
 \left(l_3 0 \Lambda' 0 | \bar l' 0\right)
 \nonumber\\
 &\quad\times
 \left(\lambda_2 0,\lambda_3\!-\!\lambda_3',0 | \lambda 0\right)
 \left(\lambda_2 0 \lambda_3' 0 | \lambda' 0\right)
 \nonumber\\
 &\quad\times
 \left(\lambda_1 0, \lambda_3\!-\!\lambda_3'', 0| \Lambda 0\right)
 \left(\lambda_1 0 \lambda_3'' 0 | \Lambda' 0\right)
 \nonumber\\
 &\quad\times
 \begin{Bmatrix}
  \lambda_3\!-\!\lambda_3' & \lambda_3' & \lambda_3 \\
  \lambda'                 & \lambda    & \lambda_2
 \end{Bmatrix}
 \begin{Bmatrix}
  \lambda_3\!-\!\lambda_3'' & \lambda_3'' & \lambda_3 \\
  \Lambda'                  & \Lambda     & \lambda_1
 \end{Bmatrix}
 \nonumber\\
 &\quad\times
 \begin{Bmatrix}
  \bar I_{12} & \bar I_{12}' & L_1 \\[3pt]
  \bar I'     & \bar I       & J
 \end{Bmatrix}
 \begin{Bmatrix}
  L_0       & L_b'\!-\!\lambda_b & l_0 \\[3pt]
  \lambda_3 & L_1                & L_2
 \end{Bmatrix}
 \begin{Bmatrix}
  1         & l_1 & l_0 \\
  \lambda_3 & L_1 & L_3
 \end{Bmatrix}
 \nonumber\\
 &\quad\times
 \begin{Bmatrix}
  \lambda_b\!-\!\lambda_b'' & \lambda_b'' & \lambda_b \\[3pt]
  L_c'\!-\!\lambda_c        & \lambda_c   & L_c'      \\[3pt]
  l_2                       & l_3         & l_1
 \end{Bmatrix}
 \begin{Bmatrix}
  \bar S_{12}' & \bar l_{12}' & \bar I_{12}' \\[3pt]
  \bar S_{12}  & \bar l_{12}  & \bar I_{12}  \\[3pt]
  L_0          & L_2          & L_1
 \end{Bmatrix}
 \begin{Bmatrix}
  \frac{1}{2} & \bar l' & \bar I' \\[3pt]
  \frac{1}{2} & \bar l  & \bar I  \\[3pt]
  1           & L_3     & L_1
 \end{Bmatrix}
 \nonumber\\
 &\quad\times
 \begin{Bmatrix}
  L_b'\!-\!\lambda_b\!-\!\lambda_b' & \lambda_b'   & L_b'\!-\!\lambda_b \\[3pt]
  \lambda                           & \lambda'     & \lambda_3          \\[3pt]
  \bar l_{12}                       & \bar l_{12}' & L_2
 \end{Bmatrix}
 \begin{Bmatrix}
  l_2     & l_3      & l_1       \\[3pt]
  \Lambda & \Lambda' & \lambda_3 \\[3pt]
  \bar l  & \bar  l' & L_3
 \end{Bmatrix},
 \label{Xcommon1}
\end{align}
where $C^{p}_{q}$ and $P_{nl}$ are the binomial coefficient
$C^{p}_{q}~\!=~\!p!/\left[\left(p-q\right)!q!\right]$ and the
momentum-space HO wave functions, respectively.
Note that the phase of $P_{nl}$ is chosen to be consistent with a
convention employed in the Fortran code \cite{Kamuntavicius01}.
The multipole-expansion function
$f_{\lambda_1\lambda_2\lambda_3}^{(L_bL_c)}$ is defined as
\begin{align}
 &f_{\lambda_1\lambda_2\lambda_3}^{(L_bL_c)}(k,k',K,K')
 \nonumber\\
 &\quad=
 \frac{\hat{\lambda}_1^2\hat{\lambda}_2^2\hat{\lambda}_3^2}{8}
 \int_{-1}^{1} \int_{-1}^{1}  \int_{-1}^{1} dw_1 dw_2 dw_3
 \nonumber\\
 &\quad\times
 P_{\lambda_1}(w_1) P_{\lambda_2}(w_2) P_{\lambda_3}(w_3)
 \nonumber\\
 &\quad\times
 \left(\left|\vect{k}-\vect{k}'\right| \left|\vect{K}-\vect{K}'\right| \right)^{-\lambda_3}
 \frac{2^{-\frac{L_b}{2}}\left(\frac{2}{3}\right)^{\frac{L_c}{2}}
   q_b^{2-L_b}q_c^{2-L_c}}{\left(q_b^2+m_\pi^2\right)\left(q_c^2+m_\pi^2\right)},
 \label{MPEfunc}
\end{align}
where $P_{\lambda_m}$ is the Legendre polynomial with
$w_m~\!=~\!\cos\theta_m~ (m=1,~2,~\textrm{or}~3)$.

\bibliographystyle{apsrev}
\bibliography{biblio.bib}

\end{document}